\documentclass[journal]{IEEEtran}
\usepackage[numbers, sort&compress, comma, square]{natbib}
\usepackage[unicode]{hyperref}  % Hyperlinks bib references.

\ifCLASSINFOpdf
  \usepackage[pdftex]{graphicx}
  \graphicspath{{./figures/}}
  \DeclareGraphicsExtensions{.pdf,.jpeg,.png}
\else
  \usepackage[dvips]{graphicx}
  \graphicspath{{./figures/}}
  \DeclareGraphicsExtensions{.eps}
\fi

\usepackage{amsmath}
\usepackage{supertabular,booktabs}
\usepackage{longtable}
\usepackage{color}
\usepackage{colortbl}
\usepackage{cleveref} % clever references

\Crefformat{figure}{#2Fig.~#1#3}
\Crefmultiformat{figure}{Figs.~#2#1#3}{ and~#2#1#3}{, #2#1#3}{ and~#2#1#3}

\usepackage{array}

\ifCLASSOPTIONcompsoc
 \usepackage[caption=false,font=normalsize,labelfont=sf,textfont=sf]{subfig}
\else
 \usepackage[caption=false,font=footnotesize]{subfig}
\fi

\ifCLASSOPTIONcaptionsoff
 \usepackage[nomarkers]{endfloat}
\let\MYoriglatexcaption\caption
\renewcommand{\caption}[2][\relax]{\MYoriglatexcaption[#2]{#2}}
\fi

\usepackage{url}

\begin{document}
%
%\title{Deep learning for low-magnitude earthquake detection on a multi-level sensor network: Application to data from Groningen, the Netherlands}

\title{Deep learning for low-magnitude earthquake detection on a multi-level sensor network}

% \author{Michael~Shell,~\IEEEmembership{Member,~IEEE,}
%         John~Doe,~\IEEEmembership{Fellow,~OSA,}
%         and~Jane~Doe,~\IEEEmembership{Life~Fellow,~IEEE}% <-this % stops a space
\author{Ahmed Shaheen,~%
        Umair~bin~Waheed,~%
        Michael Fehler,~%
        Lubos Sokol,~%
        and~Sherif Hanafy% <-this % stops a space
\thanks{Ahmed Shaheen, Umair bin Waheed, and Sherif Hanafy are with the Department of Geosciences, King Fahd University of Petroleum and Minerals, Dhahran 31261, Saudi Arabia~(e-mails: ahmed.afify.shahin@gmail.com; umair.waheed@kfupm.edu.sa; sherif.mahmoud@kfupm.edu.sa).}% <-this % stops a space
\thanks{Michael Fehler is with the Department of Earth, Atmospheric, and Planetary Sciences, Massachusetts Institute of Technology, Cambridge, MA 02139, United States~(e-mail: fehler@mit.edu).}% <-this % stops a space
\thanks{Lubos Sokol is with Seismik s.r.o., Kubi\v{s}ova~1265/8, 182~00 Praha~8, Prague, Czech Republic~(e-mail: lubos@seismik.cz).}% <-this % stops a space
%\thanks{Manuscript received ...; revised ....}
}

%\markboth{IEEE Transactions on Neural Networks and Learning Systems,~Vol.~XX, No.~X, XXXX~2021}%
%{Shaheen \MakeLowercase{\textit{et al.}}: Event detection using CNN}


%\IEEEspecialpapernotice{(Special Issue on Deep Learning for Earth and Planetary Geosciences)}

\maketitle

\begin{abstract}
Automatic detection of low-magnitude earthquakes has become an increasingly important research topic in recent years due to a sharp increase in induced seismicity around the globe. The detection of low-magnitude seismic events is essential for microseismic monitoring of hydraulic fracturing, carbon capture and storage, and geothermal operations for hazard detection and mitigation. Moreover, the detection of micro-earthquakes is crucial to understand the underlying mechanisms of larger earthquakes. Various algorithms, including deep learning methods, have been proposed over the years to detect such low-magnitude events. However, there is still a need for improving the robustness of these methods in discriminating between local sources of noise and weak seismic events. In this study, we propose a convolutional neural network (CNN) to detect seismic events from shallow borehole stations in Groningen, the Netherlands. 
We train a CNN model to detect low-magnitude earthquakes, harnessing the multi-level sensor configuration of the G-network in Groningen. Each G-network station contains four geophones at depths of 50, 100, 150, and 200 meters. Unlike prior deep learning approaches that use 3-component seismic records only at a single sensor level, we use records from the entire borehole as one training example. This allows us to train the CNN model using moveout patterns of the energy traveling across the borehole sensors to discriminate between events originating in the subsurface and local noise arriving from the surface. We compare the prediction accuracy of our trained CNN model to that of the STA/LTA and template matching algorithms on a two-month continuous record. We demonstrate that the CNN model shows significantly better performance than STA/LTA and template matching in detecting new events missing from the catalog and minimizing false detections. Moreover, we find that using the moveout feature allows us to effectively train our CNN model using only a fraction of the data that would be needed otherwise, saving plenty of manual labor in preparing training labels. The proposed approach can be easily applied to other microseismic monitoring networks with multi-level sensors.
\end{abstract}


% \fi
\IEEEpeerreviewmaketitle

% Sections
\section{introduction}
\IEEEPARstart{A}{utomatic} detection of low-magnitude earthquakes has been a longstanding research problem in seismology. Although earthquakes are known to be clustered in time and space, the underlying processes that connect one earthquake to the next are still poorly understood. The challenge comes from our limited capability to detect small earthquakes reliably~\cite{brodsky2019importance}. Given the sharp rise in induced seismicity during the past decade, the study of these low-magnitude events is critical for waste-water injection, geothermal monitoring, carbon capture and storage (CCS), and hydraulic fracturing in shale gas reservoirs. Moreover, the exponential growth of seismic data being collected, due to a huge increase in the number of recording stations over the years, makes automatic detection a necessity. Therefore, a robust and accurate detection algorithm is essential to get the full potential of the available data and to understand the subsurface processes better.

One of the earliest approaches to automate seismic event detection was the STA/LTA algorithm~\cite{allen1978automatic}. The algorithm computes the ratio of a short-term average of the amplitudes (STA) to a long-term average of the amplitudes (LTA). This ratio yields a high value when an event is present. If the ratio exceeds a preset threshold value, a trigger is activated, indicating the presence of a seismic event. The algorithm shows robustness in detecting events with high signal-to-noise ratio (SNR) data. However, the performance of STA/LTA drops significantly when used to detect events for low SNR data. For these cases, STA/LTA fails to differentiate between time-varying noises and seismic events.

To enhance the detection of events for low SNR waveforms, the template matching technique~\cite{shelly2007non} has been quite successful. Template matching cross-correlates a template waveform of a prior master seismic event to continuous seismic records. A high correlation between the waveforms of the records and those of the template indicates the presence of an event. The technique has shown to be efficient in detecting events buried under noise. However, its major drawbacks are insensitivity to events with waveforms that are dissimilar to the master event and high computational cost.

Several other techniques have been proposed over the years to tackle the problem of automatic event detection. \cite{yoon2015earthquake} proposed the FAST algorithm, which is a data mining method, to detect uncataloged events with good computational efficiency. \cite{poliannikov2018instantaneous} stacked the instantaneous phases of several traces to detect seismic events in the vicinity of a master event. They applied the known moveout of the master event on the traces before stacking. Since the instantaneous phases of noise form a Gaussian distribution, an event will appear on this stacked data as an outlier from this distribution. \cite{mukuhira2020low} used the polarization property of the P-wave to detect events in low SNR data. Machine learning techniques have also been developed in recent years to detect low-magnitude events robustly~\cite{akram2017robust,qu2018automatic}. However, there is still a need for improvement in pushing the envelope further to detect more of such small magnitude earthquakes reliably.

Deep learning has emerged as a powerful technique to tackle longstanding problems in various data-intensive fields such as computer vision and speech recognition. The ability of deep learning to recognize complex structures in high-dimensional data makes it attractive over other conventional algorithms~\cite{lecun2015deep}. In image classification, for instance, deep learning algorithms have proved significantly more accurate than any preceding technique~\cite{szegedy2015going,simonyan2014very}. Moreover, deep learning can extract features directly from the data, requiring little or no feature engineering of the raw data. 

Improved performance over conventional methods and ease of implementation have encouraged applications of deep learning on various seismological problems, including passive seismic event detection~\cite{kong2019machine}. \cite{perol2018convolutional} trained a convolutional neural network (CNN) to detect earthquakes from seismic records in Guthrie, Oklahoma. The region is known for problems with earthquakes caused by induced seismicity. The alteration of stresses in the subsurface due to waste-water injection has led to the increasing number of earthquakes in Oklahoma~\cite{keranen2013potentially}. They used around 834,000 labeled samples of events and noise to train the network. The trained network was tested on a month-long continuous seismic record. The network detected ten times more events than those present in the catalog with 94\% precision, where precision is the number of true detected events to the total number of event detections. Other studies have also shown the efficacy of deep learning in reliably detecting passive seismic events~\cite{mousavi2019cred,mousavi2020earthquake}.

Detection and study of these, predominantly low-magnitude, events caused by induced seismicity help in understanding subsurface mechanisms and assessing hazard probabilities. Another region that suffers from the problem of induced seismicity is Groningen, the Netherlands. Groningen has suffered over the past three decades from induced seismicity, caused by production from the Groningen gas field, the largest gas field in Western Europe. Continuous gas production since the 1960s led to reservoir compaction, which has been causing induced seismic events since 1991~\cite{van2015induced}.

Since the turn of the century, these events have significantly increased in both the frequency of occurrence and magnitude. These seismic events pose a danger to the lives and properties of residents in Groningen and neighboring areas. The recurrence of these events forced the Dutch government to announce the stopping of production by 2022~\cite{reuters.USKCN1VV1KE}, causing huge financial loss to the country and operators. Therefore, understanding the underlying mechanisms of these earthquakes is vital to minimize further human and financial losses in Groningen and, more importantly, to prevent problems from escalating in other regions of the world. The first step to study these mechanisms is having a robust and accurate algorithm to detect low-magnitude events.

In this study, we use data recorded by the shallow borehole network in Groningen, known as the G-network. The network consists of borehole stations that contain multi-level geophones at increasing depths of 50 m. Since CNNs are well-known for identifying spatial features in data, we use the moveout pattern of the energy traveling across the borehole sensors at a single station as the main distinguishing factor between events originating in the subsurface and noise coming from the surface. This is often the most challenging problem in detecting low-magnitude earthquakes. On a single sensor, earthquake energy coming from below or energy coming from a local noise source on the surface may look the same, leading to the difficulty in classification. In contrast, the two sources of energy will show opposite moveout patterns on a multi-level sensor network. Just as a human interpreter could visually separate true events from coherent noise by picking the moveout, i.e., up-down (noise) or down-up (event), we train a CNN for this classification task. %This is motivated by the remarkable ability of CNNs to recognize spatial variations within data. 
Recognizing these moveout patterns increases the network’s resolution in detecting events of lower magnitudes and in minimizing false detections. 

We test the trained network on two months (December, 2017 – January, 2018) of continuous records at five stations near Zeerijp – the site of a major event ($M_L$ = 3.4) that occurred on January 8, 2018. We compare the findings of the network to the outputs of the STA/LTA and template matching algorithms on the same two-month continuous record. We demonstrate that despite being trained on a relatively small amount of data, the network shows significantly better performance than STA/LTA and template matching in detecting new events missing from the catalog and minimizing false detections. In total, we manage to increase the number of detected events by about 100\% than those in the original catalog. 

The rest of the paper is organized as follows: We begin by explaining the methodology used to train and test the network. This is followed by the results obtained using the proposed method and its comparison with conventional techniques. Finally, we discuss the key findings of this study and its implications.

\section{Methodology}

In this section, we begin by outlining details of the seismic network used in the study. Next, we illustrate the pre-processing steps applied to the seismic records before training the CNN. Then, we shed some light on CNN’s key features, followed by a description of the network’s architecture. Lastly, we describe how the network’s performance on continuous data is evaluated and compared to other methods.

\subsection{Training Data Selection and Pre-processing}

\begin{figure}[htb]
	\centering
	\includegraphics[width=0.55\linewidth]{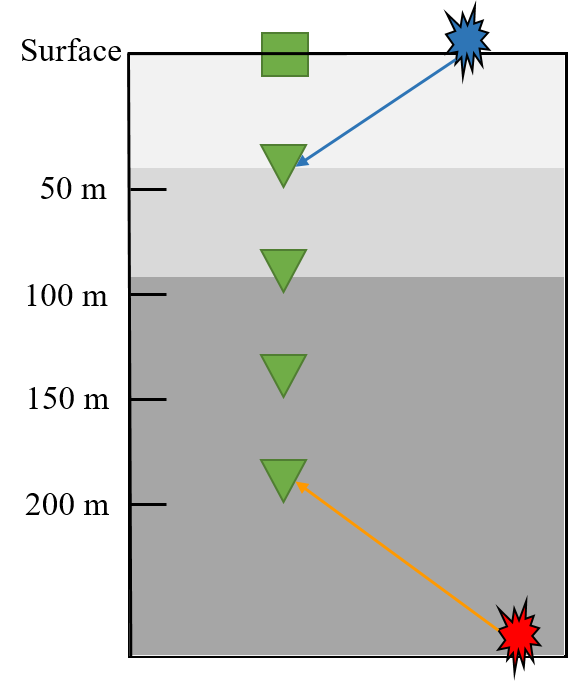}
	\caption{A typical G-network station consisting of an accelerometer at the surface (green square) and four geophones (green triangles). This setup allows the differentiation between genuine seismic events (red) and noise coming from the surface (blue) by observing the moveout pattern across the sensors. \textit{Drawn after~\cite{dost2017development}}.}
	\label{G-station}
\end{figure}

The increase in the number and strength of induced seismicity connected to the Groningen gas field resulted in the densification of the monitoring network over the years. In particular, after the largest event in the region ($M_L$ = 3.6) that occurred in 2012 near Huizinge, a total of 70 stations were added to cover the Groningen gas field with an average spacing of 4-5 km. Together, these 70 stations form the G-network that is comprised of shallow borehole stations~\cite{dost2017development}. Each borehole station contains an accelerometer at the surface and four velocity geophone sensors (3-component). These velocity sensors are located at depths of 50~m, 100~m, 150~m, and 200~m, respectively, from the surface. In this study, we use records from the four velocity geophones for training and testing the CNN model. Figure~\ref{G-station} shows the configuration of a typical G-network station.

\begin{figure}[htb]
	\centering
	\includegraphics[width=0.85\linewidth,keepaspectratio]{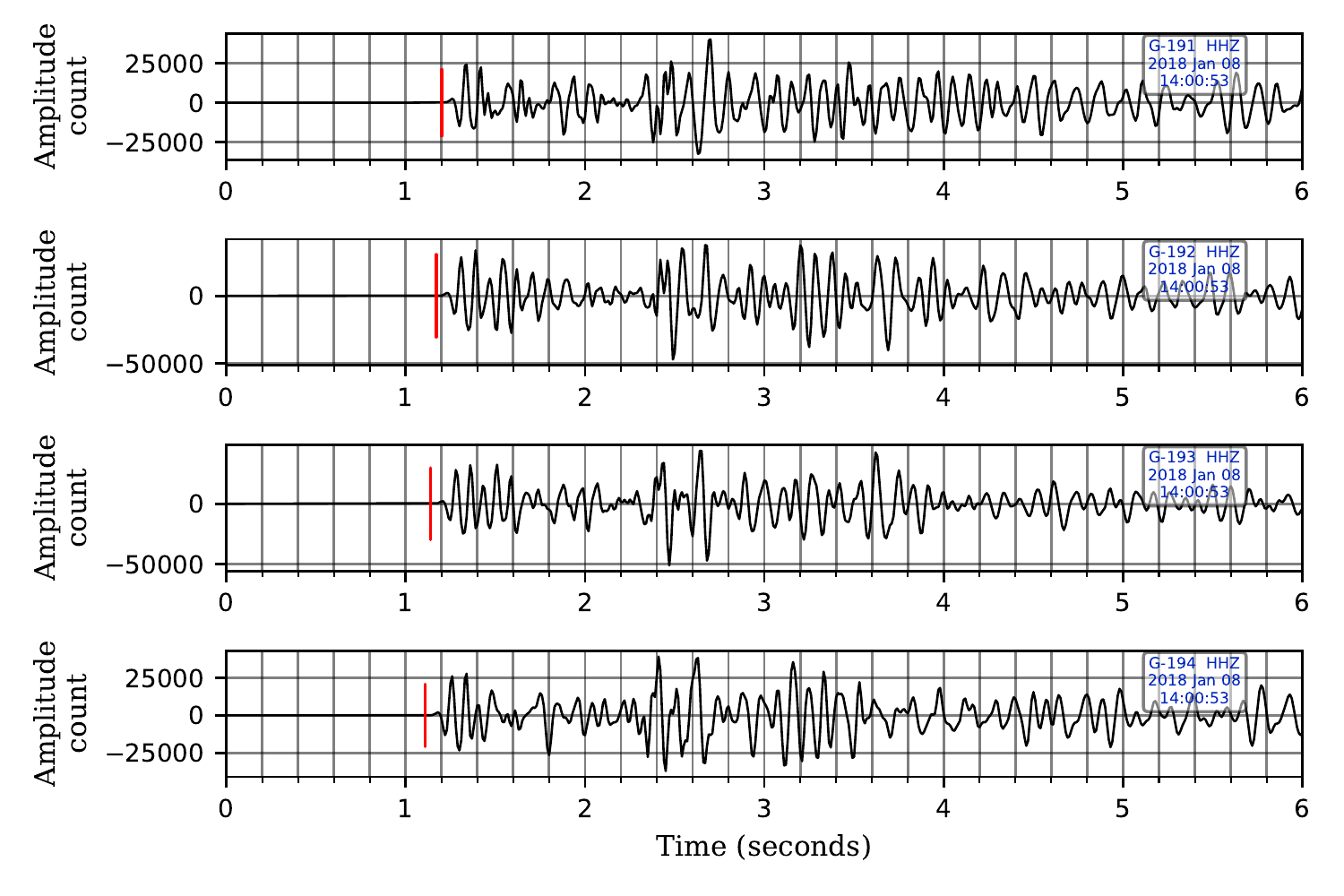}
	\caption{The moveout pattern of a genuine seismic event observed across the four geophone levels. The red vertical lines denote the first-arrival times of the seismic wave at each sensor. The wave arrives first at the deepest sensor and propagates upwards. A bandpass filter between 5-25 Hz has been applied to the records. Note that only vertical components are shown here for illustration.}
	\label{normal moveout}
\end{figure}

\begin{figure}[htb]
	\centering
	\includegraphics[width=.85\linewidth,height=\textheight,keepaspectratio]{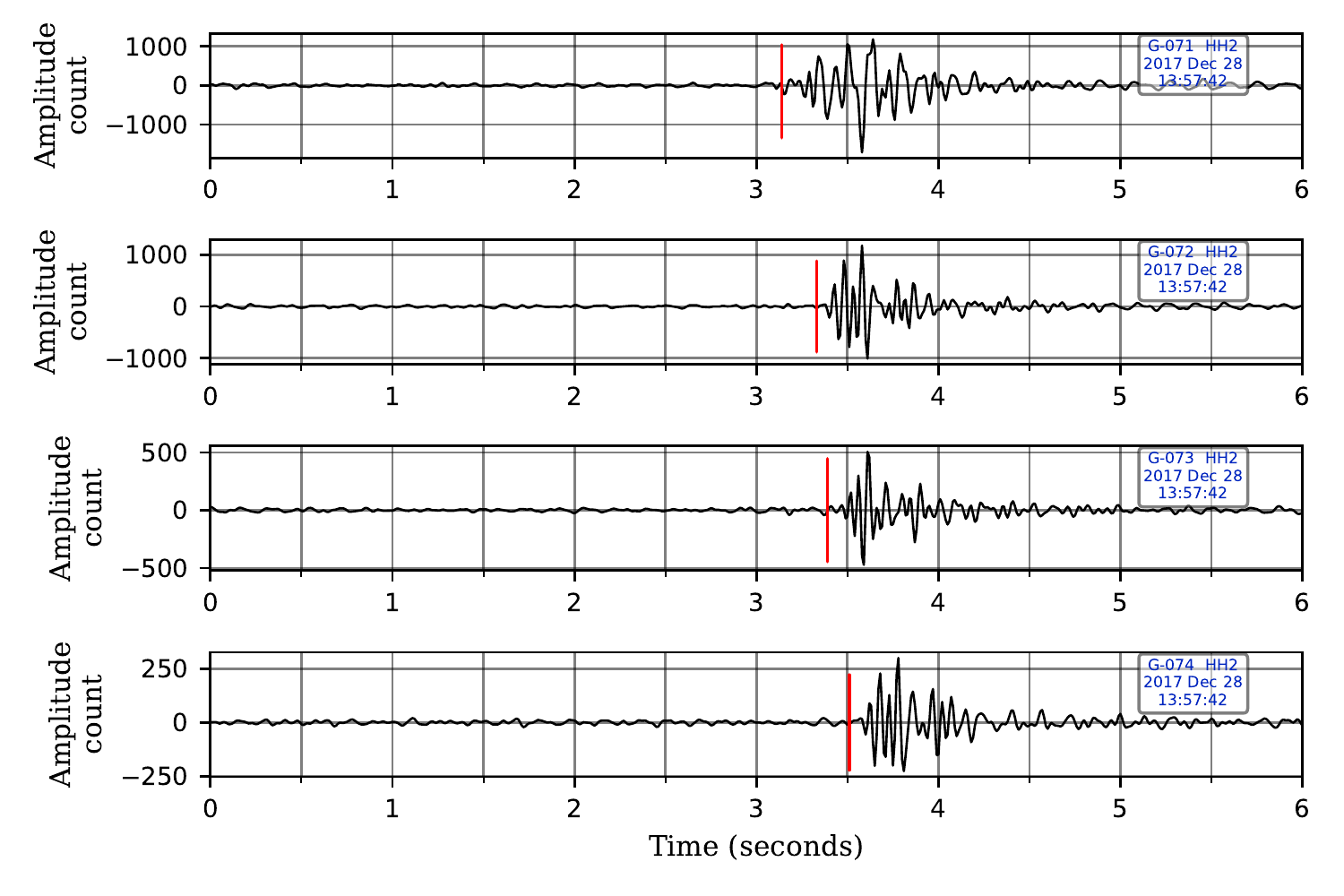}
	\caption{The moveout pattern of noise coming from the surface observed across the four geophone levels. The red vertical lines denote the first-arrival times of the seismic wave at each sensor. The wave arrives first at the shallowest sensor and propagates downwards. A bandpass filter between 5-25 Hz has been applied to the records. Note that only one of the horizontal components is shown here for illustration.}
	\label{reverse moveout}
\end{figure}

We harness this borehole station setup to improve the robustness of our CNN model in distinguishing events from local sources of noise. True seismic events will arrive first at the deepest sensor, while noise coming from the surface arrives first at the shallowest sensor. This results in a moveout pattern – bottom-up for true events in contrast to up-down for coherent noise sources on the surface. We teach a CNN model, using carefully selected waveforms, to distinguish between such moveout differences and improve classification accuracy. Classification algorithms that use data only at a single sensor are likely to suffer from inaccurate predictions in the presence of strong coherent noise as it cannot effectively distinguish coherent noise from true events if observed only at a single sensor level. Figure~\ref{normal moveout} shows the moveout pattern of a genuine seismic event across the four velocity geophones of a G-network station, and Figure~\ref{reverse moveout} shows an example of the moveout of noise coming from above.

Seismic records of the G-network are made available online by the Royal Netherlands Meteorological Institute at its seismic and acoustic data portal~\cite{knmi1993netherlands}. In particular, we are interested in the time period around the large event ($M_L$=3.4) in the Zeerijp region that occurred on January 8, 2018. Therefore, we consider data from all the 47 events listed in the KNMI catalog between October 1, 2017, till February 28, 2018, for training a CNN model. For each of the 47 listed events, we retrieve four-minute long records at all the 70  G-network stations starting from three minutes before the listed origin time until one minute after the origin time. The time-windows for the event class are taken from the interval after the origin time, and the time-windows for the noise class are taken from the interval before it. Moreover, additional time-windows containing coherent noise coming from the surface were searched for and added as noise time-windows. The length of these event and noise windows are taken to be 30~s long, which we empirically found to be the most suitable length for most events in the region.

All time-windows undergo pre-processing before being used to train the CNN model. The time windows are detrended, demeaned, and bandpass filtered between 5-25 Hz. We found this band of frequencies to be most suitable in removing considerable ambient noise without much effect on the signal of interest, making the events more obvious to detect. Furthermore, we down-sample the data from 200 Hz to 100 Hz to reduce the data size by a factor of 2. This helps improve the efficiency of the training process as fewer time samples are needed to be processed by the CNN model. Finally, these time-windows are manually verified to remove any window that does not fit its corresponding label. 

\begin{figure}[htb]
	\centering
	\includegraphics[width=.85\linewidth,height=\textheight,keepaspectratio]{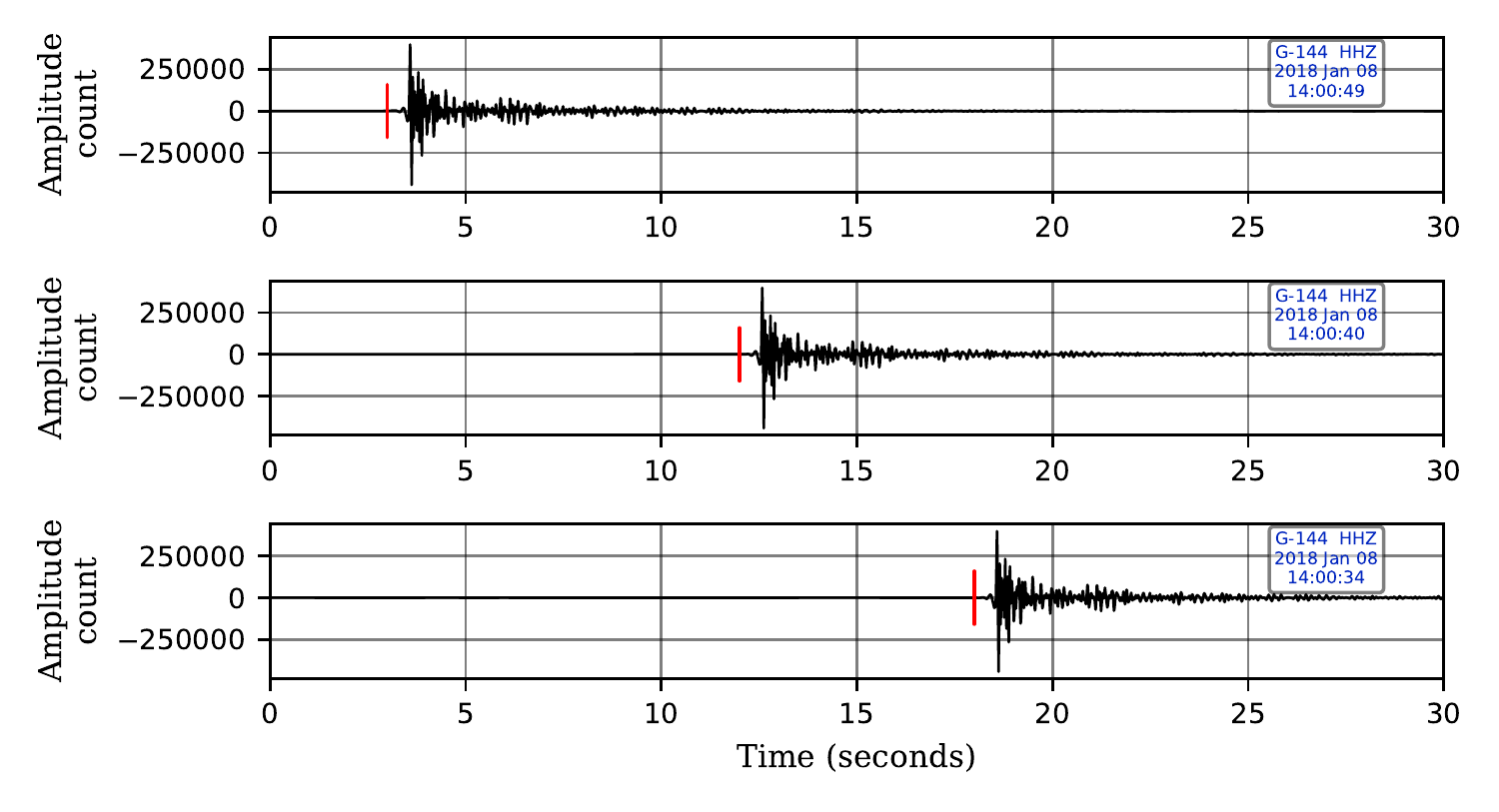}
	\caption{Three examples of the shifts done on an event time-window. Each  event window is taken 17 times with 17 different placements of the arrival time to the window start: 3 sec shift (top), 12 sec (middle), 18 sec (bottom). Red lines indicate the first-arrival times. \label{shifting}}
\end{figure}

Furthermore, to help the CNN model generalize better to variations in the test data, we augment our training event windows by adding time shifts such that the first-break time is varied with respect to the start of the window. This step is necessary to avoid any potential network bias towards events that have first-breaks at particular time-samples within an event window. For each event window, we generate an additional sixteen windows by moving the window back and forth with a randomly generated time-shift. Figure~\ref{shifting} shows a signal time-window with three of the seventeen total windows coming from a single original event window. Without data augmentation, this bias may affect the network’s performance when testing on a continuous record since, in a continuous record, the signal can arrive at any time sample in a given window. The same data augmentation technique is applied to noise windows as well. Finally, we normalize the waveforms from each of the four sensor levels individually to have a maximum amplitude of unity. 

%\begin{figure}[!ht]
%	\centering

%	\subfloat{\includegraphics[width=\columnwidth]{./figures/Station-G-144-shift_3.pdf}}\\[-0.35cm]
%	\subfloat{\includegraphics[width=\columnwidth]{./figures/Station-G-144-shift_12.pdf}}\\[-0.35cm]
%	\subfloat{\includegraphics[width=\columnwidth]{./figures/Station-G-144-shift_18.pdf}}

%	\caption{Three examples of the shifts done on an event time-window. Each  event window is taken 17 times with 17 different placements of the arrival time to the window start: 3 sec shift (top), 12 sec (middle), 18 sec (bottom).\label{shifting} {\color{red} Figure 4 still is not like the other two figures. It has xlabels on all levels etc. Please fix this} {\color{blue}  done}}
%	\end{figure}

The output after the afore-mentioned pre-processing workflow is a 30 second time-window labeled into either signal or noise. The shape of each training data is a 3D matrix whose dimensions are: 4 (number of velocity geophones) $\times$ 3001 (time-window length $\times$ sampling rate) $\times$ 3 (the 3-components of each sensor). This is an analogous setup to the one used in the classification of color images, where the first two dimensions are the image length and width, and the third dimension represents the three color channels (red, green, and blue). The total number of labeled examples is 67,847: 21,624 belonging to the event class and 46,223 belonging to the noise class. These labeled examples are split according to the following distribution: 60\% for training, 20\% for validation, and 20\% for testing.

\subsection{CNN Architecture}
\label{cnn-architecture}

Convolutional neural networks (CNNs) are inspired by studies on the visual cortex of cats and monkeys~\cite{hubel1959receptive,hubel1968receptive}. A CNN is a neural network with multiple hidden layers that uses convolutions to generate features hierarchically. Its main advantage over a feed-forward neural network is that it does not require tedious feature selection and engineering of raw data before feeding to the network. Moreover, a CNN can work with inputs in their original shape (whether it is 2D, 3D, or even 4D) without the need to flatten them into a 1D vector as required by feed-forward neural networks. This helps in locating spatial variations in multi-dimensional data better. Furthermore, due to the filters in its convolutional layers, a CNN shows higher efficiency in detecting local features within data compared to a feed-forward neural network.

A convolutional layer is a key part of CNN, where feature extraction takes place. The layer extracts features from data by convolving filters of small dimensions with the input data. Then, the result of this convolution process is passed through a non-linear activation function. The output after the activation function is referred to as a feature map. A feature map carries important information from the input data after being filtered. A convolution layer can produce tens of feature maps, each containing a certain trait of the data that helps in efficiently performing the task at hand, be it a classification or a regression problem. 

The convolutional layer produces a huge amount of data that need to be scaled down for computational efficiency without losing important information before proceeding further. Therefore, each convolutional layer is followed by a max-pooling layer. This layer reduces data dimension by taking the maximum value of neighboring inputs in the feature maps. Getting the maximum ensures that the most important values are passed on. The reduced output of the pooling layer is passed to the subsequent convolutional layer to obtain more abstract features, followed by another pooling layer, and so on. This sequence of convolutional and pooling layers continues as many times as necessary to build highly complex features. Then, the output of the final pooling layer is flattened and passed into fully connected layers before passing on to the final layer that outputs the result. The convolutional filter coefficients and network weights are learned through the process of backpropagation~\cite{rumelhart1985learning}.

\begin{figure}[!ht]
	\centering
	\includegraphics[height=0.9\textheight,keepaspectratio]{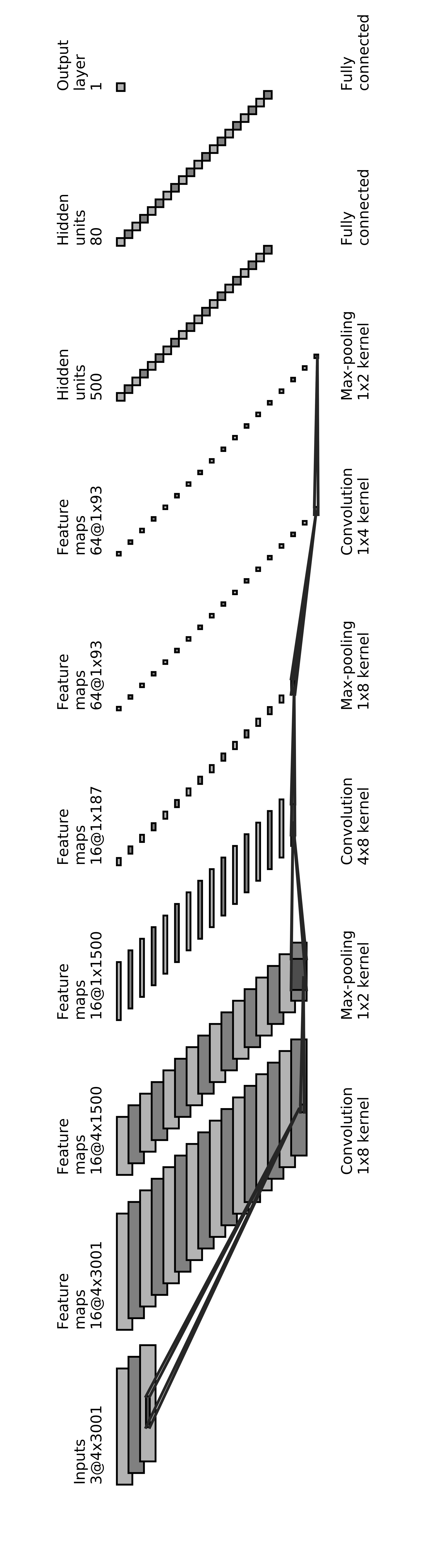}
	\caption{The proposed CNN architecture starting with 3 convolutional layers, each followed by a max-pooling layer. The output of the final pooling layer is flattened and passed into two fully connected layers before outputting the result: event (1) or noise (0).}
	\label{my CNN}
\end{figure}

Our proposed CNN architecture consists of 3 convolutional layers. Each of them is followed by a max-pooling layer. The output of the last pooling layer is flattened and passed into two fully connected layers, followed by the output layer, which consists of a single neuron outputting whether the input is an event (1) or noise (0). Table~\ref{CNN-arch} shows a summary of the network architecture, and Figure~\ref{my CNN} shows the network pictorially. This architecture was chosen empirically after testing several architectures. %and parameters. %The chosen parameters gave the best performance.

The activation function used in all layers, except for the final layer, is the rectified linear unit function (ReLU). The final layer uses the sigmoid activation function. We use the Adam optimizer, and the loss function used to compute the error is the binary cross-entropy. The network is implemented using the Tensorflow library~\cite{tensorflow2015-whitepaper} in Python. The network's performance is evaluated by calculating the accuracy of its predictions on all datasets (training, validation, \& testing). The network stops training when the validation accuracy ceases to improve for eight consecutive epochs. The network then gets the weights of the epoch with the best validation accuracy. 

\begin{table}
	\caption{A detailed summary of the CNN architecture, showing the number and dimensions of filters and neurons at each layer of the CNN.}
	\label{CNN-arch}
	\centering
	\begin{tabular}{|>{\centering}p{0.18\linewidth}|>{\centering}p{0.16\linewidth}|>{\centering}p{0.13\linewidth}|>{\centering}p{0.13\linewidth}|c|}
		\hline
		Layer &
		\begin{tabular}[c]{@{}c@{}}Input \\ Dimension\end{tabular} &
		\begin{tabular}[c]{@{}c@{}}Number of \\ Filters\end{tabular} &
		\begin{tabular}[c]{@{}c@{}}Kernel \\ Dimension\end{tabular} &
		\begin{tabular}[c]{@{}c@{}}Output \\ Dimension\end{tabular} \\ \hline
		Conv Layer 1 & 4$\times$3001$\times$3  & 16  & 1$\times$8 & 4$\times$3001$\times$16 \\ \hline
		Pool Layer 1 & 4$\times$3001$\times$16 & -   & 1$\times$2 & 4$\times$1500$\times$16 \\ \hline
		Conv Layer 2 & 4$\times$1500$\times$16 & 16  & 4$\times$8 & 1$\times$1500$\times$16 \\ \hline
		Pool Layer 2 & 1$\times$1500$\times$16 & -   & 1$\times$8 & 1$\times$187$\times$16  \\ \hline
		Conv Layer 3 & 1$\times$187$\times$16  & 64  & 1$\times$4 & 1$\times$187$\times$64  \\ \hline
		Pool Layer 3 & 1$\times$187$\times$64  & -   & 1$\times$2 & 1$\times$93$\times$64   \\ \hline
		Flatten      & 1$\times$93$\times$64   & -   & -          & 5952$\times$1           \\ \hline
		Dense 1      & 5952$\times$1           & 500 & -          & 500$\times$1            \\ \hline
		Dense 2      & 500$\times$1            & 80  & -          & 80$\times$1             \\ \hline
		Output Layer & 80$\times$1             & 1   & -          & 1                       \\ \hline
	\end{tabular}
\end{table}

\begin{figure}[htb]
	\centering
	\includegraphics[width=.85\linewidth]{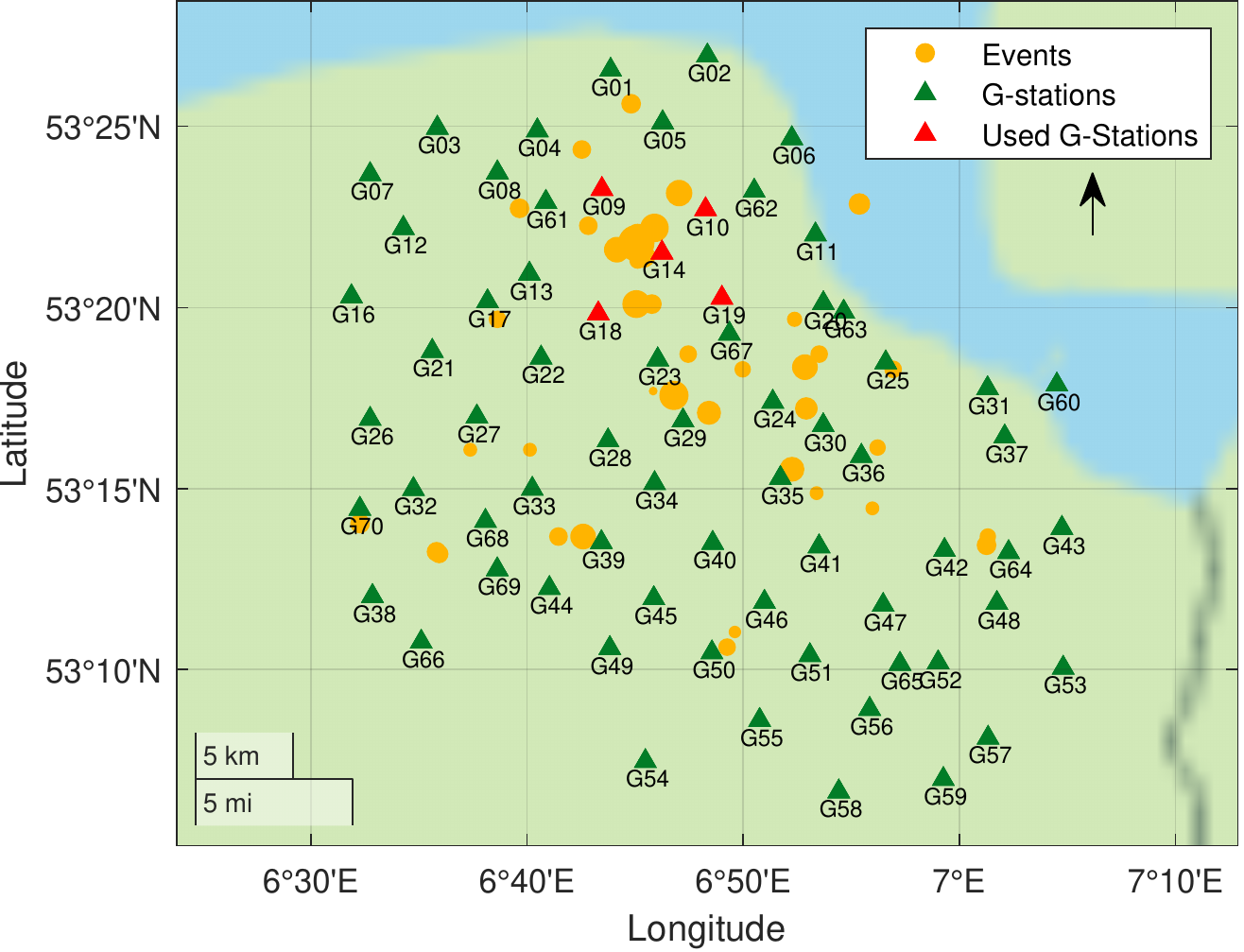}
	\caption{Map of the G-network stations (green and red) and the 23 cataloged events (orange) that occurred during December, 2017 and January, 2018. The selected five stations (shown in red) are used to test the trained CNN model on a two-month continuous record.}
	\label{testing basemap}
\end{figure}

\subsection{Performance Evaluation on Continuous Data}

To evaluate performance of the trained model on continuous data, the network is tested on recordings from five stations (G09, G10, G14, G18, G19) during the months of December, 2017 and January, 2018. In these two months, 23 events are recorded in the catalog, including a major event near Zeerijp ($M_L$=3.4) on January 8. The chosen five stations are the closest to the epicenter of this major event. Figure~\ref{testing basemap} shows locations of the cataloged events that occurred during the two-month period and the G-network stations. We highlight, in red, the five stations whose records are used for performance evaluation.

The two-month record is cut into overlapping 30-second time-windows. Each subsequent 30-second window starts 10 seconds after the start time of the preceding time-window. This ensures that the network does not miss any event that may get split between two consecutive windows. Every time-window undergoes the same pre-processing steps that were used for the training data (detrending, demeaning, bandpass filtering, downsampling by a factor of 2, and amplitude normalization to unity). We require a given time-window to be flagged as an event class on at least two of the five stations for it to be classified as an event. This helps us reduce the possibility of false alarms and ensures the robustness of the CNN predictions. We use the available event catalog and manual verification to analyze the accuracy of these predictions.

For comparison, we run STA/LTA with several detection thresholds on the same continuous data coming from the selected five stations. The four velocity geophones at each station are stacked to enhance the SNR. This stacking is done to improve the STA/LTA’s detection accuracy. Similar to CNN, to be classified as an event, we require at least two of the five stations to cross a given threshold. The findings of the STA/LTA are classified into true events, uncertain events, and false detections. The uncertain class refers to time-windows that are categorized as events by STA/LTA but are unverifiable visually due to low SNR data. 

Template matching is also run on the same stations for the two-month period. It uses the 23 cataloged events during the two-month period as master events to search for undetected events. Template matching results are also compared to the CNN’s results in the following section.

\section{Results}

In this section, we provide training details of the CNN architecture, outlined in Section~\ref{cnn-architecture}, and its performance on training, validation, and test datasets. This is followed by the analysis of the trained network's performance on a two-month continuous record and its comparison with the STA/LTA and template matching algorithms.

\begin{figure}

	\centering
	
	\begin{tabular}{c c}
    \subfloat{\includegraphics[scale=0.25]{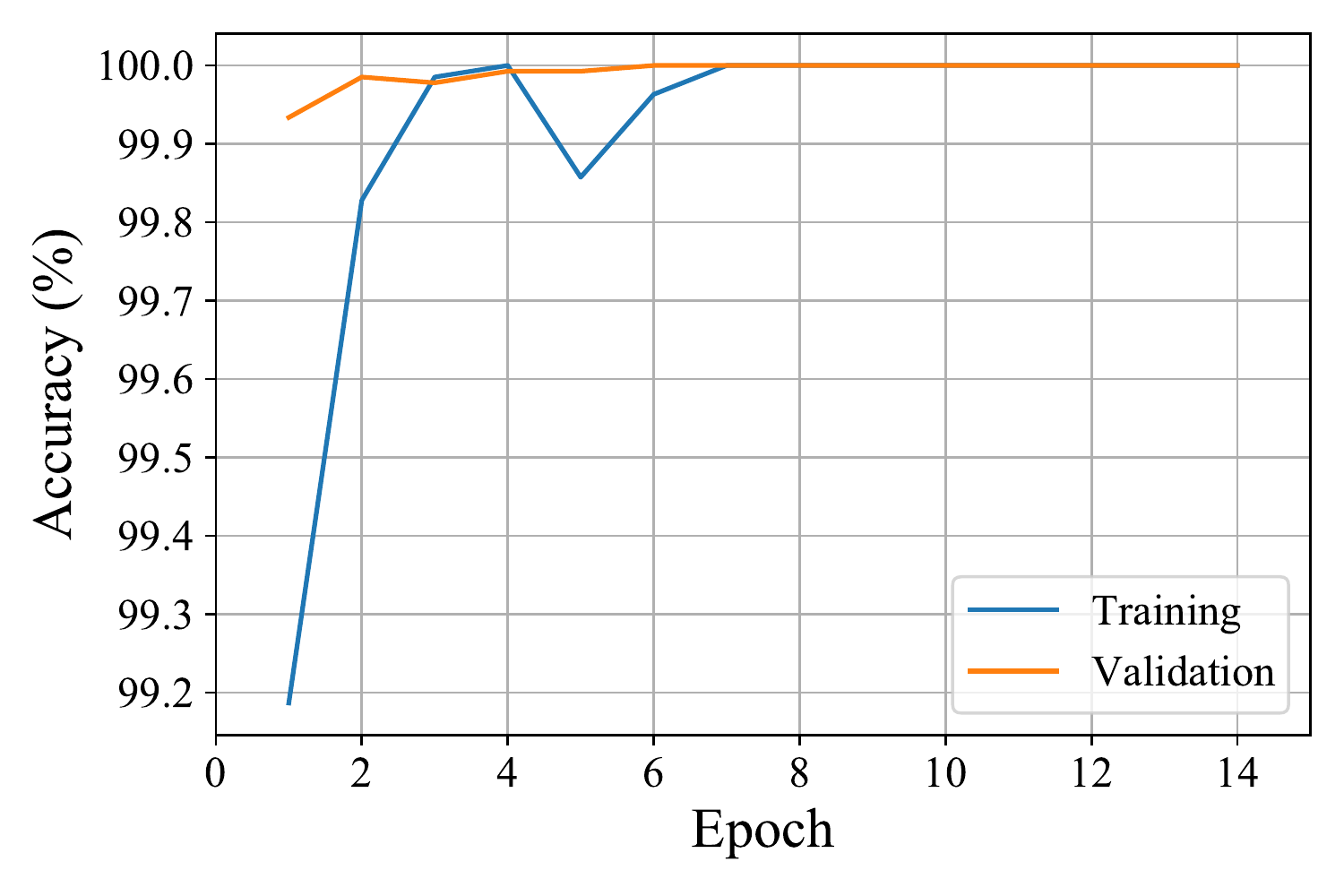}}
   
    \subfloat{\includegraphics[scale=0.25]{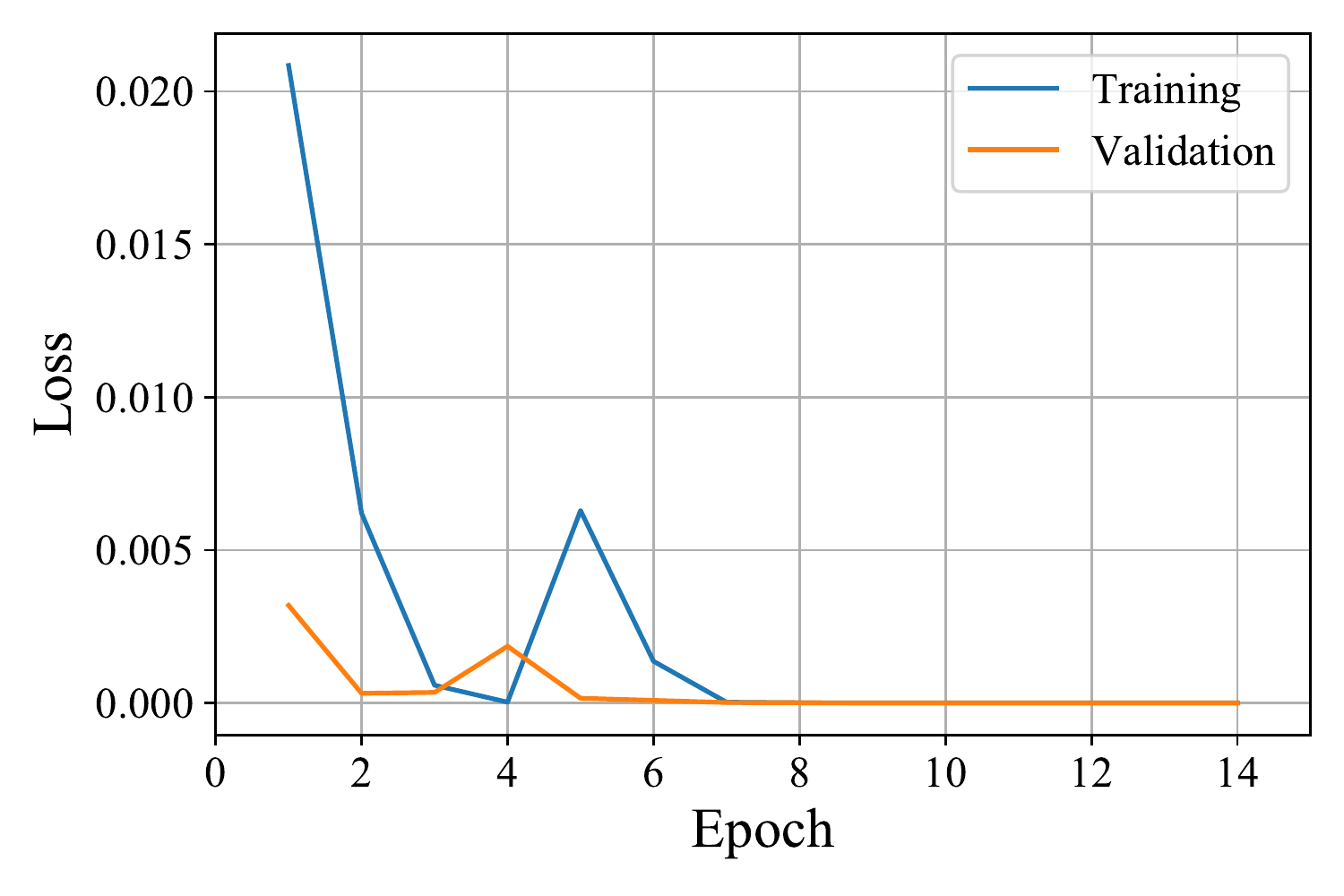}}
    \\
    \end{tabular}
	\caption{Accuracy and loss of training and validation datasets vs. epochs. The network converges after a few epochs to the optimum solution.}
	\label{network training}
	
\end{figure}

The network is trained with an Adam optimizer using mini-batch optimization with a batch size of 64. The training stops after 14 epochs as validation accuracy ceases to improve for eight consecutive epochs. Figure~\ref{network training} shows the accuracy and the loss of the training and validation datasets versus the training epochs. We observe high accuracy for both the training and validation datasets even after the first epoch as we use a small batch size to help with the convergence speed of the network. We obtain 100\% accuracy on the training, validation, and testing datasets. 

%The network’s performance on training, validation and testing datasets shows that the network is not overfitting the training data. Over-fitting is when a neural network performs with high accuracy on training dataset but performs badly on testing dataset. When over-fitting occurs, it indicates that the network fails to generalize its findings to unseen data. The 100\% accuracy on testing dataset shows that the network is generalizing its findings and is not over-fitting on training labels. 

%\begin{figure}
%	\centering
%	\includegraphics[width=\linewidth,keepaspectratio]{./figures/Loss.pdf}
%	\caption{Loss of training and validation datasets vs epochs.}
%	\label{network training:loss}
%\end{figure}

\begin{figure}[!ht]
	\centering

	\subfloat{\includegraphics[width=0.85\columnwidth]{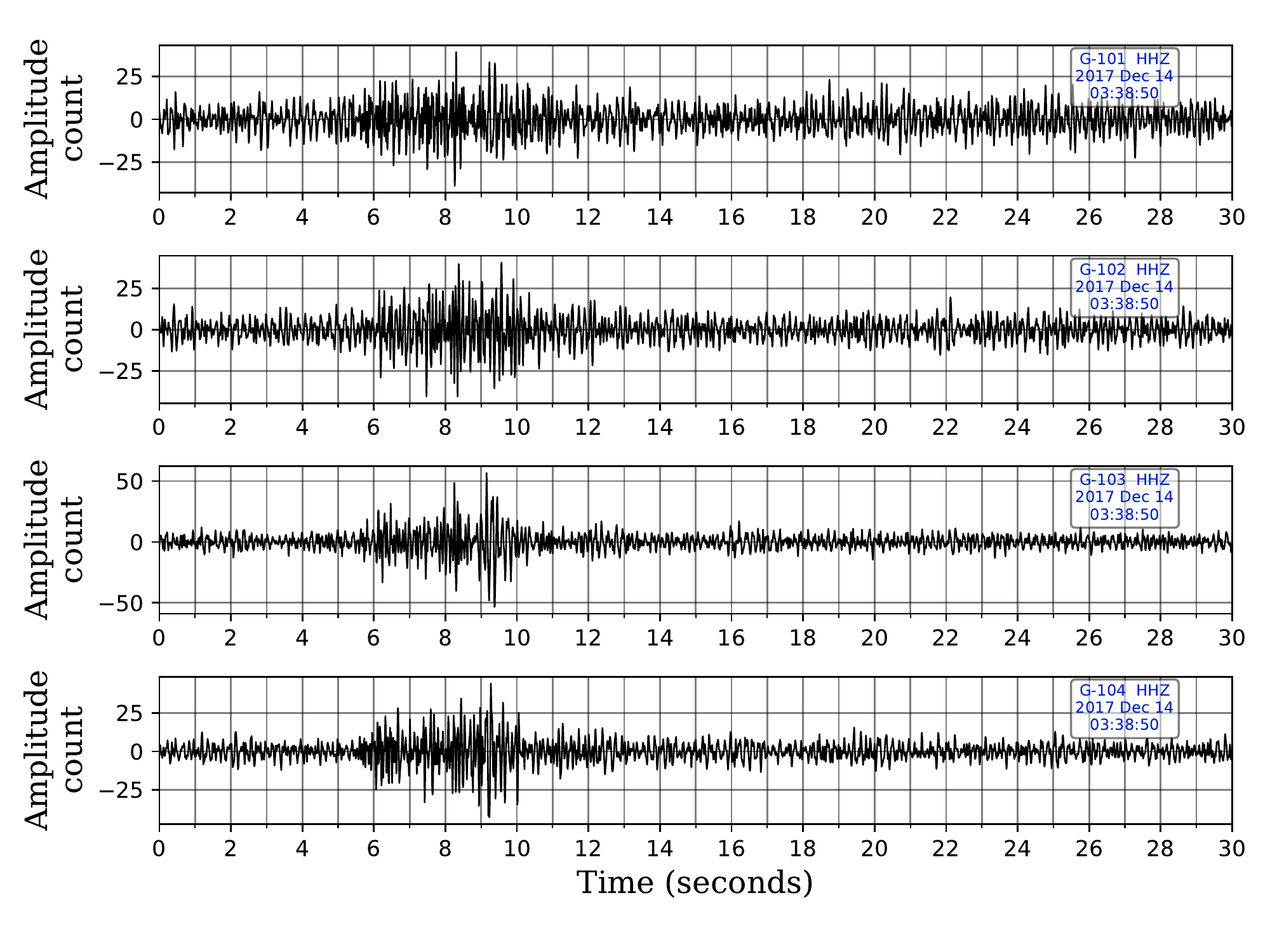}}
	
	\qquad
	\subfloat{\includegraphics[width=0.85\columnwidth]{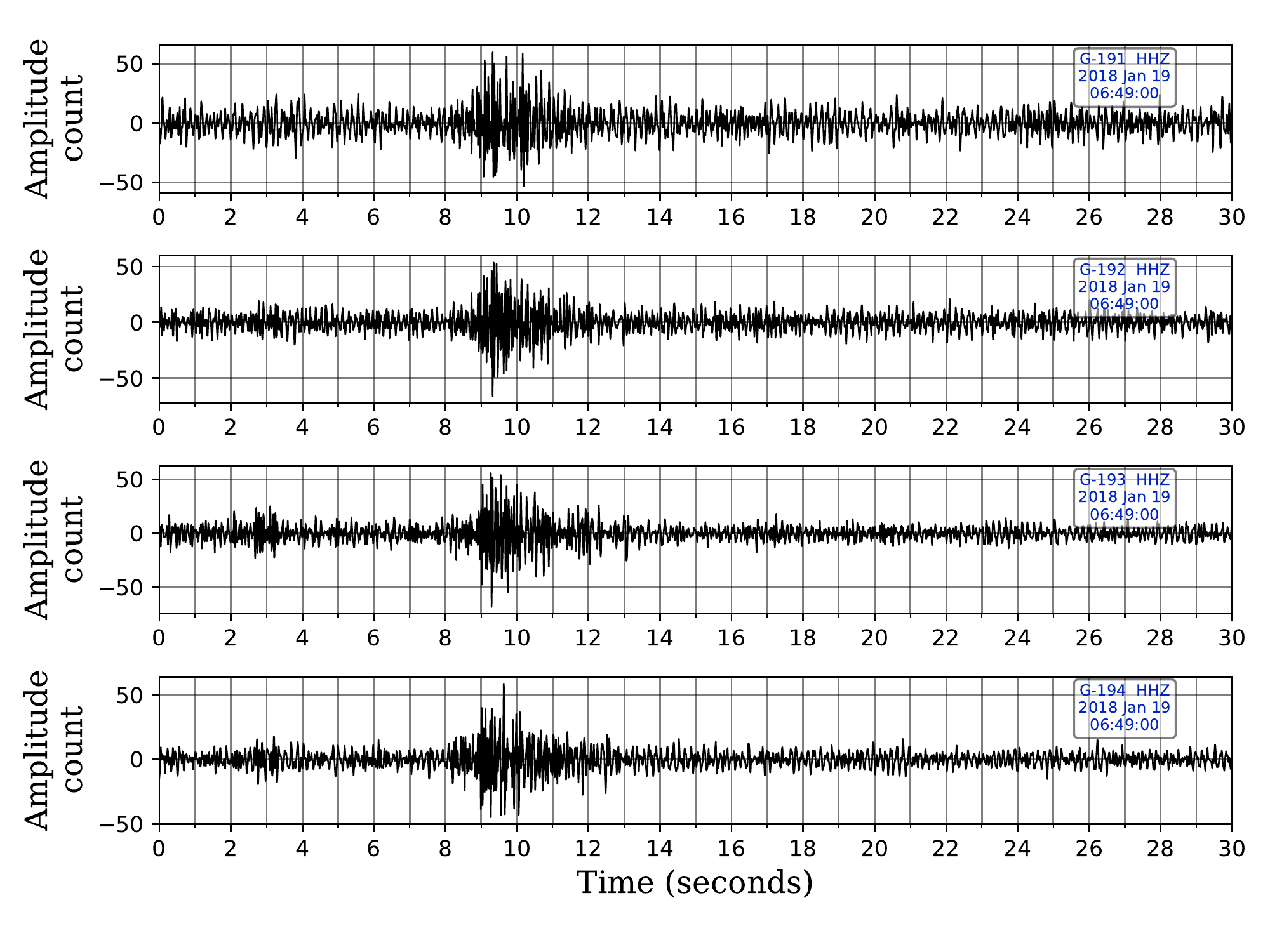}}
	
	\caption{Z-component waveforms at the four sensor levels for stations G10 and G19 from two of the uncataloged events picked by our trained CNN model. These two events were neither picked by STA/LTA nor template matching. Picking of these two events, despite their low-magnitude and relatively low SNR data, shows the efficacy of the trained CNN model. 
	\label{CNN events}}
\end{figure}

The network is then tested on two months of continuous data at the five stations near Zeerijp, indicated with red triangles in Figure~\ref{testing basemap}. The trained CNN model flags a total of 45 time-windows as events. Upon manual verification and comparison with the existing catalog, we find 40 of those to be true events while 5 were found to be false detections. It picks 20 of the 23 cataloged events that were reported during the two-month period. The three unpicked cataloged events are actually very low-magnitude events ($M_L$ = 0.4, 0.5 \& 0.7) with epicenters considerably far away from the selected stations. Upon manually looking at the five selected stations at these reported time-windows, we find the signals to be undetectable visually as they were buried under noise. Missing these events using the selected five stations is understandable since our CNN model was trained using events that we can visually confirm, and therefore, it is unable to identify such events. However, the trained network picks 20 more low-magnitude events that were not cataloged. Figure~\ref{CNN events} shows the Z-component records of two of these uncataloged events with low SNR data that were picked by the CNN. Appendix~\ref{app:CNN events} details the 45 picked time-windows and their classifications.

%% Events Figures

\begin{figure}[htb]
	\centering
	\includegraphics[width=0.76\linewidth]{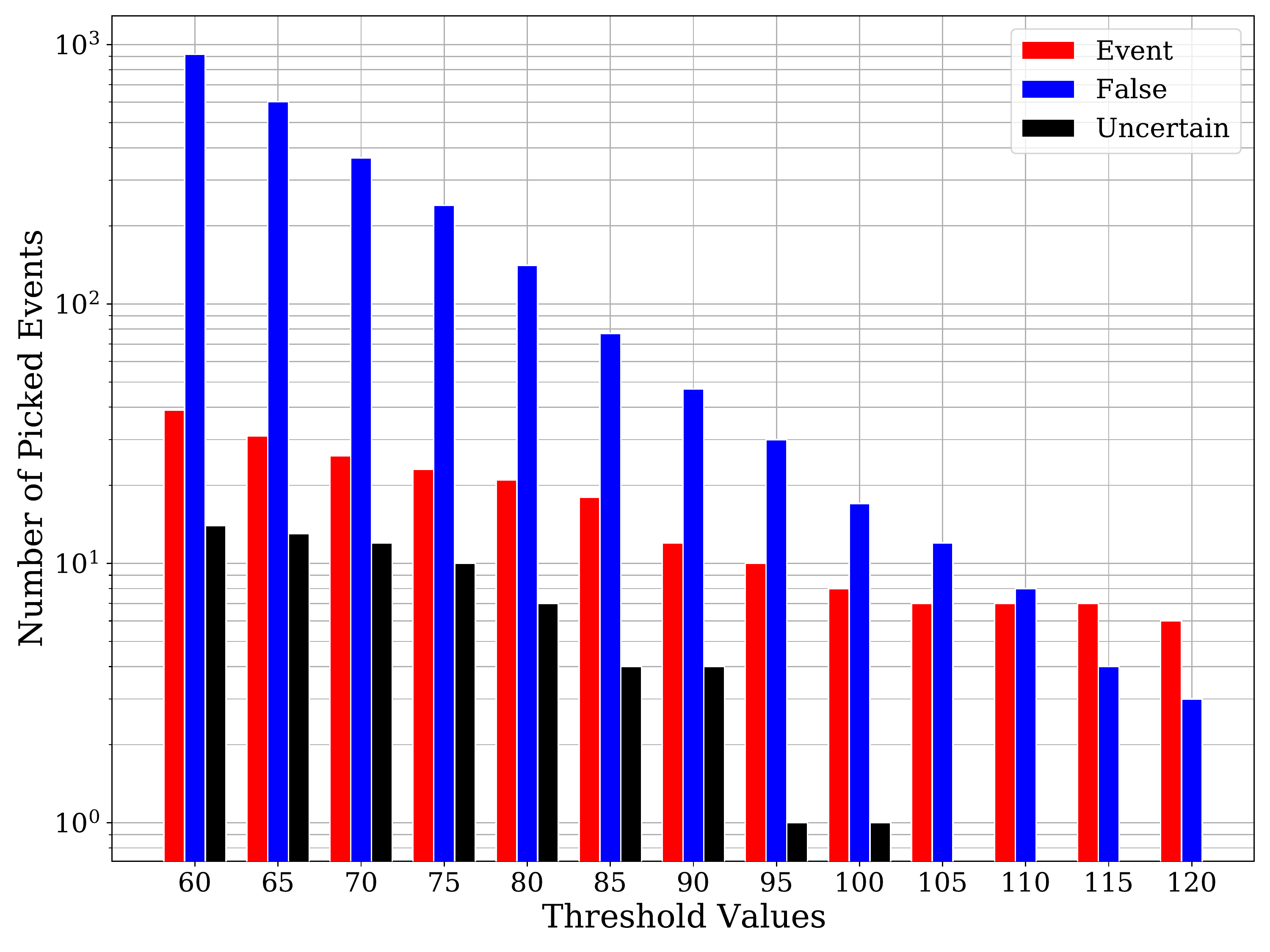}
	\caption{Barplot of STA/LTA detections with their classification at different threshold values. Using low threshold values STA/LTA picks numerous false detections, while stricter threshold values result in missing lots of true events.}
	\label{sta pickings}
\end{figure}

Moving on to the comparison with conventional techniques, it is well-known that the sensitivity of STA/LTA predictions is highly dependent on the chosen threshold value. Low threshold values lead to a lower risk of missing true events at the cost of getting a higher number of false detections. On the contrary, a threshold value that is too high may avoid those false detections but leads to missing true events. Therefore, we apply STA/LTA on the two-month continuous record for a range of threshold values. Figure~\ref{sta pickings} shows the number of true events, uncertain events (not verifiable manually), and false detections for a range of threshold values. For the threshold value of 60, we get 39 true events, 14 uncertain events, and 917 false detections. Despite using a low threshold value, STA/LTA was able to pick only 14 out of the 23 cataloged events. Besides the cataloged events, STA/LTA also picks 8 true events that were also identified by CNN and 16 weak events that were not picked by the CNN model. However, the extremely high number of false detections renders this approach to be unacceptable. 

\begin{figure}[htb]
	\centering
	\includegraphics[width=0.76\linewidth]{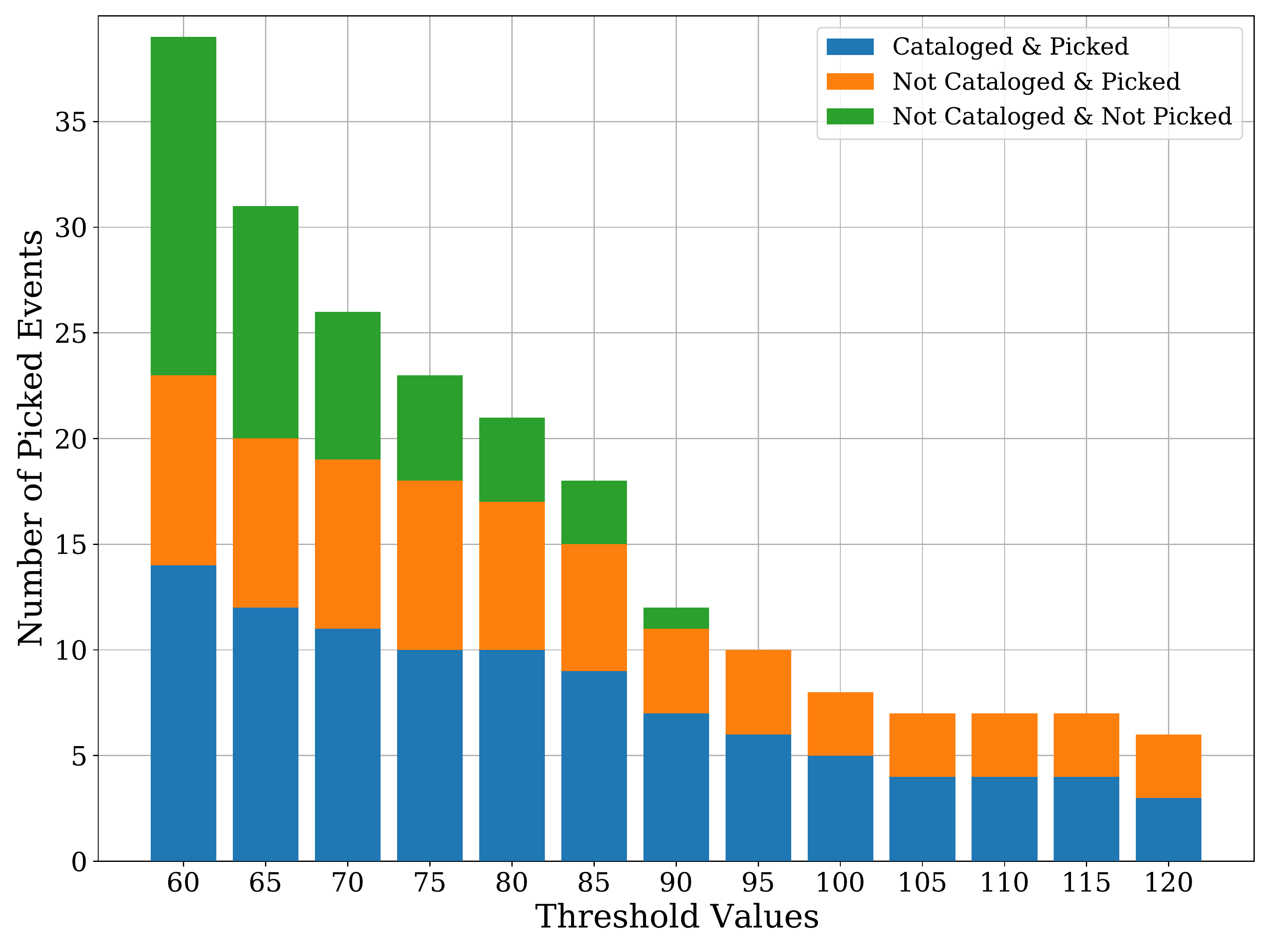}
	\caption{A barplot showing the detailed classes of events picked by STA/LTA for different threshold values. The categorization shows how many of these events have been cataloged/uncataloged and whether they were picked by our CNN model or not.
	}
	\label{sta events}
\end{figure}

On the other end of the spectrum, i.e., for a threshold value of 120, we manage to reduce the number of false detections to only 3 but the true picks are also substantially reduced to only 6, missing many cataloged events. Figure~\ref{sta events} shows the detailed classification of the STA/LTA event detections at each threshold value, highlighting cataloged or uncataloged events and whether these events were picked by our CNN model or not.

%STA/LTA is tried using ascending threshold values to compare its finding with the network’s findings. The STA/LTA either picks numerous false alarms on using low thresholds or misses a lot of events with more strict thresholds. Figure~\ref{sta pickings} shows how many true events, uncertain events, and false alarms detected at each threshold value. On the least used threshold (threshold 60), the network picks 39 events, 14 uncertain events, and 917 false alarms. It picks only 14 of the 23 cataloged events. This means it misses 6 cataloged events that are picked by the network. Beside 14 cataloged events, the STA/LTA picks 25 more events, 8 of them are picked the CNN while the rest are weak new detections. Figure~\ref{sta events} shows the detailed classes of the events detections (cataloged or not cataloged and picked by the CNN or not) at each threshold.

To compare the CNN and STA/LTA detection performance, we compute the precision value of the CNN and STA/LTA detections. Precision is the ratio of the true positive findings to the summation of the true and false positives. For STA/LTA, we exclude the uncertain events from computing the precision, which is computed using the following formula:

\begin{equation*}
	Precision = \dfrac{True\:Events}{True\:Events  +   False\:Events}
\end{equation*}

\begin{figure}[htb]
	\centering
	\includegraphics[width=0.76\linewidth,keepaspectratio]{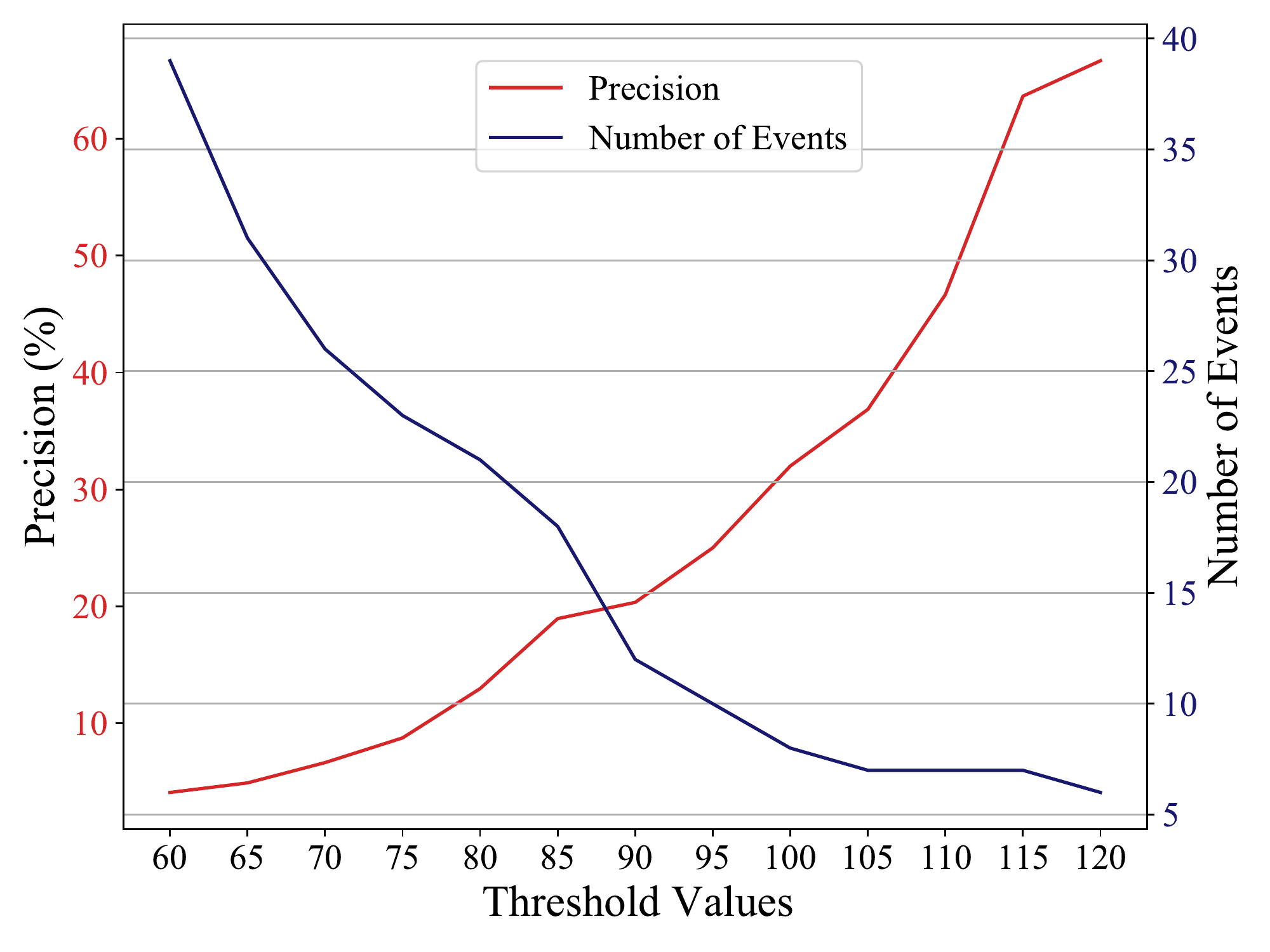}
	\caption{Precision of STA/LTA along with the number of the detected events at each threshold value. Using a higher threshold value increases the precision but at the expense of reducing the number of detected true events. Even with the highest threshold value, the STA/LTA precision (67\%) remains less than the CNN precision (88.9\%), and it picks only 6 events for this case compared to 40 events picked by the CNN model.
	}
	\label{sta precision}
\end{figure}

We evaluate the precision of the CNN model on the two-month dataset to be 88.9\%, while the STA/LTA detection precision does not exceed 67\% on any of the threshold values tried. STA/LTA only reaches the 67\% precision value with the highest threshold used, where it detects only 6 events. Figure~\ref{sta precision} shows the precision of STA/LTA together with the number of the detected events at each threshold value. Since STA/LTA relies on sudden changes in amplitude to differentiate an event from noise, the under-performance of the algorithm in detecting low-magnitude earthquakes is understandable. Since weak events and coherent noise yield similar waveform signatures, the STA/LTA algorithm ends up either detecting many false alarms or missing true events, depending on the chosen threshold value. Moreover, as opposed to the CNN model, it can neither recognize the polarization of the seismic events on the 3-component data nor the difference between moveout patterns across multi-level geophones.

%STA/LTA detects all these false alarms despite stacking of the four levels and using the two stations condition to improve STA/LTA’s performance. STA/LTA detects events by detecting sudden changes in amplitudes that accompany seismic events. When aiming to detect such low-magnitude events, both weak events and random noise will show similar change in amplitude causing many false detections. Moreover, STA/LTA can neither recognize the polarization of the seismic events on the 3-component data nor the difference between moveout patterns across multi-level geophones.

\begin{figure}[htb]
	\centering
	\includegraphics[width=0.85\linewidth,keepaspectratio]{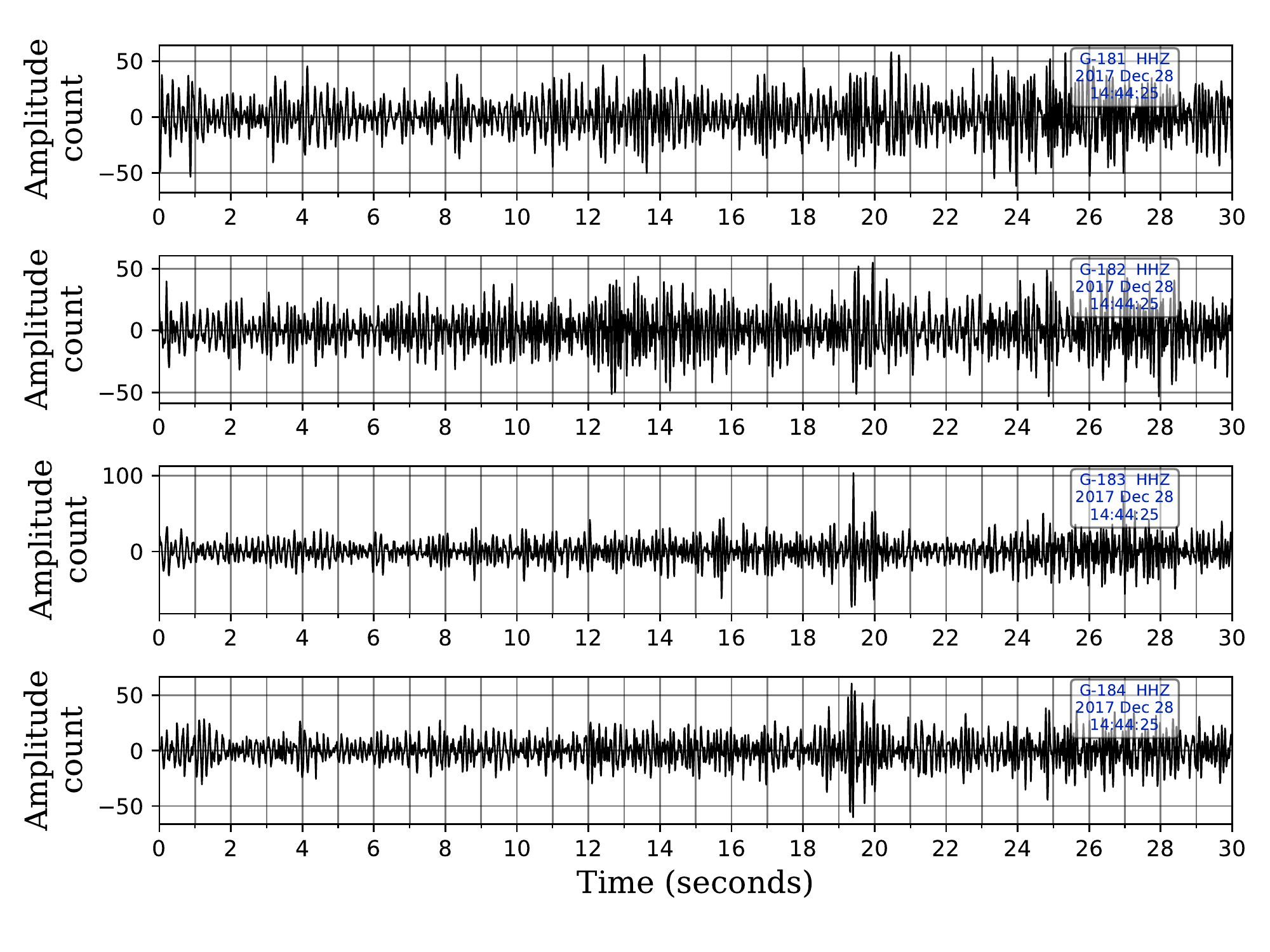}
	\caption{An event detected by template matching and missed by our CNN model at $14:44:32.929$ hours on December 28, 2017. The event is buried under noise and the network was not trained on such event waveforms. Only the vertical waveforms are shown for illustration.}
	\label{TM event}
\end{figure}

We also analyze the detection performance using template matching. Since it uses all the 23 cataloged events as master events, it finds them all during the two-month period, which is unsurprising. In addition, template matching detects 10 uncataloged events. This is half the number of uncataloged detections by the CNN model. Appendix~\ref{app:template} lists the uncataloged events captured by template matching and compares them with the uncataloged events detected by the CNN model. We note that the CNN detects 8 of the 10 events that are identified by template matching as well, i.e., template matching picks 2 events that are not picked by the CNN model. These two events are at $06:19:50.809$ hours on December 1, 2017, and $14:44:32.929$ hours on December 28, 2017. Investigation of these two events reveals that the first event is not picked by the CNN model because of the two-station requirement to flag an event. This event is visually detectable on only one station and is correctly picked at this station by the CNN. However, it is not flagged as an event because it was buried under noise at other stations. The second event, shown in Figure~\ref{TM event}, is buried under noise at all five stations. Since the CNN model was trained only on events that were visually detectable, it is understandable that it missed these events.

\section{Discussion and conclusions}

We developed a CNN-based algorithm for detecting low-magnitude events in Groningen using a multi-level sensor network. We find the performance of the trained CNN model to be considerably superior in comparison with STA/LTA and template matching. We test the performance of these methods on a two-month dataset around the relatively large $M_L=3.4$ event on January 8, 2018. While many other deep learning based techniques have also been developed in recent times, our proposed methodology takes advantage of the multi-level sensor network in discriminating between true events arriving from the subsurface and local noise coming from the surface by training the network to distinguish between moveout patterns at the borehole sensors. Moreover, unlike template matching, the method is not sensitive to only the waveform signatures of the master events. Furthermore, once the CNN model is trained, it can provide detection results in real-time, which is also an important consideration for hazard mitigation in microseismic monitoring.

Prior deep learning studies on the problem of seismic event detection~\cite{perol2018convolutional,ross2018p,mousavi2019cred} used hundreds of thousands to millions of labels to train their networks. In this study, we show that remarkably high detection performance can be achieved by using a relatively small amount of data. We used only 3,991 labels that were further augmented to create 17 times more training data. This can be attributed to the idea of using the moveout at the multi-level sensor network as a discriminating feature, allowing us to gain maximum leverage from such a relatively small amount of data, which also saves us time in manual labeling and verification of the training data.

However, one downside of our proposed CNN model is its lack of ability to detect events buried under noise. This characteristic is due to the data used to train the network. We only used data that a human interpreter could visually confirm as events and, therefore, it did not include events that were masked by noise. One approach to overcome this limitation is to prepare training data using template matching since it has the capability to detect events even in a negative SNR regime. This also highlights an important direction for deep learning practitioners where existing tools and methodologies can be used to further improve the capabilities of these modern learning algorithms instead of viewing them solely as competing approaches. 

\section{Acknowledgments}

We acknowledge the support by the College of Petroleum Engineering and Geosciences at KFUPM. We extend gratitude to Prof. Stewart Greenhalgh for useful discussions.

\newpage

\appendices

\section{Events picked by the trained CNN model}
\label{app:CNN events}
%\tablefirsthead{\toprule	\textbf{Events}                                       & \textbf{CNN detected}  & \textbf{Template Matching detected}  }

%\tablefirsthead{\toprule	\textbf{Events}                                       & \textbf{CNN detected}  & \textbf{Template Matching detected}  }

%% Table 1: CNN Events

%\tablecaption{CNN Detections. Calaloged events are highlighted in green color, uncataloged in yellow and false alarms in red. \label{tab:CNN events}}\\ 

\tablefirsthead{%
\hline
\multicolumn{1}{|c}{\textbf{Time window start}}                                 & \multicolumn{1}{|c}{\begin{tabular}[c]{@{}c@{}}\textbf{Number of}\\\textbf{stations} \end{tabular}} & \multicolumn{1}{|c}{\textbf{Class}}  & \multicolumn{1}{|c|}{\textbf{Comments}} \\
\hline}
	
\tablehead{%
\hline
\multicolumn{1}{|c}{\textbf{Time window start}}                                 & \multicolumn{1}{|c}{\begin{tabular}[c]{@{}c@{}}\textbf{Number of stations}\\\textbf{that picked the event} \end{tabular}} & \multicolumn{1}{|c}{\textbf{Class}}  & \multicolumn{1}{|c|}{\textbf{Comments}} \\
\hline}

\tabletail{	\hline \multicolumn{4}{|r|}{{\small\sl continued on next page}} \\ \hline}

\tablelasttail{\hline}

\tablecaption{CNN detections with cataloged events highlighted in green, uncataloged in yellow, and false alarms in red.\label{tab:CNN events}}

\begin{supertabular}{|c|c|c|c|}
	\rowcolor[rgb]{0.776,0.937,0.808} 2017-12-01 11:33:30 & 5                                                                                                      & Event           & Cataloged           \\ 
	\hline
	\rowcolor[rgb]{0.776,0.937,0.808} 2017-12-01 21:05:40 & 5                                                                                                      & Event           & Cataloged           \\ 
	\hline
	\rowcolor[rgb]{0.776,0.937,0.808} 2017-12-02 09:00:30 & 5                                                                                                      & Event           & Cataloged           \\ 
	\hline
	\rowcolor[rgb]{1,0.922,0.612} 2017-12-02 11:23:20     & 2                                                                                                      & Event           & Not Cataloged       \\ 
	\hline
	\rowcolor[rgb]{1,0.78,0.808} 2017-12-03 16:52:50      & 2                                                                                                      & False Alarm     & -                   \\ 
	\hline
	\rowcolor[rgb]{0.776,0.937,0.808} 2017-12-06 23:29:00 & 3                                                                                                      & Event           & Cataloged           \\ 
	\hline
	\rowcolor[rgb]{1,0.78,0.808} 2017-12-10 15:55:30      & 2                                                                                                      & False Alarm     & -                   \\ 
	\hline
	\rowcolor[rgb]{0.776,0.937,0.808} 2017-12-10 16:48:30 & 5                                                                                                      & Event           & Cataloged           \\ 
	\hline
	\rowcolor[rgb]{1,0.922,0.612} 2017-12-13 00:43:50     & 3                                                                                                      & Event           & Not Cataloged       \\ 
	\hline
	\rowcolor[rgb]{1,0.922,0.612} 2017-12-14 03:38:50     & 4                                                                                                      & Event           & Not Cataloged       \\ 
	\hline
	\rowcolor[rgb]{0.776,0.937,0.808} 2017-12-15 20:55:40 & 5                                                                                                      & Event           & Cataloged           \\ 
	\hline
	\rowcolor[rgb]{1,0.922,0.612} 2017-12-16 02:17:50     & 2                                                                                                      & Event           & Not Cataloged       \\ 
	\hline
	\rowcolor[rgb]{1,0.922,0.612} 2017-12-17 07:01:30     & 5                                                                                                      & Event           & Not Cataloged       \\ 
	\hline
	\rowcolor[rgb]{1,0.922,0.612} 2017-12-20 18:26:40     & 2                                                                                                      & Event           & Not Cataloged       \\ 
	\hline
	\rowcolor[rgb]{0.776,0.937,0.808} 2017-12-21 19:39:00 & 4                                                                                                      & Event           & Cataloged           \\ 
	\hline
	\rowcolor[rgb]{0.776,0.937,0.808} 2017-12-22 19:40:20 & 5                                                                                                      & Event           & Cataloged           \\ 
	\hline
	\rowcolor[rgb]{0.776,0.937,0.808} 2017-12-22 20:06:10 & 5                                                                                                      & Event           & Cataloged           \\ 
	\hline
	\rowcolor[rgb]{1,0.922,0.612} 2017-12-22 23:41:00     & 5                                                                                                      & Event           & Not Cataloged       \\ 
	\hline
	\rowcolor[rgb]{1,0.922,0.612} 2017-12-24 07:05:10     & 4                                                                                                      & Event           & Not Cataloged       \\ 
	\hline
	\rowcolor[rgb]{0.776,0.937,0.808} 2017-12-24 17:49:50 & 3                                                                                                      & Event           & Cataloged           \\ 
	\hline
	\rowcolor[rgb]{0.776,0.937,0.808} 2017-12-25 12:52:10 & 3                                                                                                      & Event           & Cataloged           \\ 
	\hline
	\rowcolor[rgb]{0.776,0.937,0.808} 2017-12-28 14:00:30 & 4                                                                                                      & Event           & Cataloged           \\ 
	\hline
	\rowcolor[rgb]{1,0.922,0.612} 2017-12-28 18:02:30     & 4                                                                                                      & Event           & Not Cataloged       \\ 
	\hline
	\rowcolor[rgb]{1,0.922,0.612} 2017-12-29 19:59:50     & 5                                                                                                      & Event           & Not Cataloged       \\ 
	\hline
	\rowcolor[rgb]{1,0.922,0.612} 2017-12-29 20:53:30     & 2                                                                                                      & Event           & Not Cataloged       \\ 
	\hline
	\rowcolor[rgb]{0.776,0.937,0.808} 2017-12-29 23:15:40 & 5                                                                                                      & Event           & Cataloged           \\ 
	\hline
	\rowcolor[rgb]{1,0.78,0.808} 2017-12-31 02:52:10      & 2                                                                                                      & False Alarm     & -                   \\ 
	\hline
	\rowcolor[rgb]{1,0.78,0.808} 2017-12-31 03:03:20      & 2                                                                                                      & False Alarm     & -                   \\ 
	\hline
	\rowcolor[rgb]{0.776,0.937,0.808} 2018-01-01 14:46:50 & 4                                                                                                      & Event           & Cataloged           \\ 
	\hline
	\rowcolor[rgb]{1,0.922,0.612} 2018-01-02 09:34:30     & 4                                                                                                      & Event           & Not Cataloged       \\ 
	\hline
	\rowcolor[rgb]{0.776,0.937,0.808} 2018-01-08 14:00:50 & 5                                                                                                      & Event           & Cataloged           \\ 
	\hline
	\rowcolor[rgb]{1,0.922,0.612} 2018-01-08 16:58:10     & 4                                                                                                      & Event           & Not Cataloged       \\ 
	\hline
	\rowcolor[rgb]{0.776,0.937,0.808} 2018-01-09 15:46:40 & 5                                                                                                      & Event           & Cataloged           \\ 
	\hline
	\rowcolor[rgb]{1,0.922,0.612} 2018-01-12 16:20:10     & 3                                                                                                      & Event           & Not Cataloged       \\ 
	\hline
	\rowcolor[rgb]{1,0.922,0.612} 2018-01-13 10:29:30     & 4                                                                                                      & Event           & Not Cataloged       \\ 
	\hline
	\rowcolor[rgb]{1,0.922,0.612} 2018-01-13 10:30:00     & 3                                                                                                      & Event           & Not Cataloged       \\ 
	\hline
	\rowcolor[rgb]{0.776,0.937,0.808} 2018-01-17 04:37:40 & 4                                                                                                      & Event           & Cataloged           \\ 
	\hline
	\rowcolor[rgb]{1,0.922,0.612} 2018-01-19 06:49:00     & 2                                                                                                      & Event           & Not Cataloged       \\ 
	\hline
	\rowcolor[rgb]{0.776,0.937,0.808} 2018-01-20 08:19:20 & 2                                                                                                      & Event           & Cataloged           \\ 
	\hline
	\rowcolor[rgb]{1,0.922,0.612} 2018-01-21 06:06:20     & 5                                                                                                      & Event           & Not Cataloged       \\ 
	\hline
	\rowcolor[rgb]{0.776,0.937,0.808} 2018-01-22 07:23:10 & 3                                                                                                      & Event           & Cataloged           \\ 
	\hline
	\rowcolor[rgb]{1,0.922,0.612} 2018-01-24 17:04:10     & 3                                                                                                      & Event           & Not Cataloged       \\ 
	\hline
	\rowcolor[rgb]{1,0.78,0.808} 2018-01-27 12:54:20      & 2                                                                                                      & False Alarm     & -                   \\ 
	\hline
	\rowcolor[rgb]{1,0.922,0.612} 2018-01-29 02:57:10     & 2                                                                                                      & Event           & Not Cataloged       \\ 
	\hline
	\rowcolor[rgb]{0.776,0.937,0.808} 2018-01-31 13:11:00 & 5                                                                                                      & Event           & Cataloged           \\
	\hline
\end{supertabular}

% Table 2 Template matching Vs CNN events.
\newpage
\section{Comparison of uncataloged detections between CNN \& template matching }

\label{app:template}

\tablefirsthead{%
\hline
\multicolumn{1}{|c|}{\textbf{Time window start}}                                       & \multicolumn{1}{|c|}{\textbf{CNN}}  & 
\multicolumn{1}{|c|}{\textbf{Template Matching}}\\  
\hline}

\tablehead{%
\hline

\multicolumn{1}{|c|}{\textbf{Time window start}}                                       & \multicolumn{1}{|c|}{\textbf{CNN}}  & 
\multicolumn{1}{|c|}{\textbf{Template Matching}}\\  
\hline}
\tabletail{%
\hline

\multicolumn{3}{|r|}{\small\sl continued on next page}\\
\hline}
\tablelasttail{\hline}
\tablecaption{Summary of the uncataloged events picked by template matching and the trained CNN model. Events picked by both template matching and CNN are highlighted in green. Events picked by template matching only are highlighted in yellow, while events picked by CNN only are highlighted in blue.
\label{tab:template}}
\begin{supertabular}{|c|c|c|}

	%
	%\caption{Template matching picked events \& uncataloged CNN pickings. Events picked by template matching and CNN are highlighted in green. Events picked by template matching only are highlighted in yellow and events picked by CNN only are highlighted in blue.\label{tab:template}}\\ 
	%\hline
	%\textbf{Events}                                       & \textbf{CNN detected}  & \textbf{Template Matching detected}   \endfirsthead 

	%\caption[]{Template matching picked events \& uncataloged CNN pickings. Events picked by template matching and CNN are highlighted in green. Events picked by template matching only are highlighted in yellow and events picked by CNN only are highlighted in blue.}\\ 
	%\hline
	%\textbf{Events}                                       & \textbf{CNN detected}  & \textbf{Template Matching detected}   \endhead
	
	%\hline \multicolumn{3}{|r|}{{Continued on next page}} \\ \hline
	%\endfoot
	
	%\hline \hline
	%\endlastfoot

	\hline
	\rowcolor[rgb]{1,0.902,0.6} 2017-12-01 06:19:50       & No                     & Yes                                   \\ 
	\hline
	\rowcolor[rgb]{0.608,0.761,0.902} 2017-12-02 11:23:20 & Yes                    & No                                    \\ 
	\hline
	\rowcolor[rgb]{0.663,0.816,0.557} 2017-12-13 00:43:55 & Yes                    & Yes                                   \\ 
	\hline
	\rowcolor[rgb]{0.608,0.761,0.902} 2017-12-14 03:38:50 & Yes                    & No                                    \\ 
	\hline
	\rowcolor[rgb]{0.663,0.816,0.557} 2017-12-16 02:17:54 & Yes                    & Yes                                   \\ 
	\hline
	\rowcolor[rgb]{0.608,0.761,0.902} 2017-12-17 07:01:30 & Yes                    & No                                    \\ 
	\hline
	\rowcolor[rgb]{0.608,0.761,0.902} 2017-12-20 18:26:40 & Yes                    & No                                    \\ 
	\hline
	\rowcolor[rgb]{0.663,0.816,0.557} 2017-12-22 23:41:05 & Yes                    & Yes                                   \\ 
	\hline
	\rowcolor[rgb]{0.663,0.816,0.557} 2017-12-24 07:05:17 & Yes                    & Yes                                   \\ 
	\hline
	\rowcolor[rgb]{1,0.902,0.6} 2017-12-28 14:44:32       & No                     & Yes                                   \\ 
	\hline
	\rowcolor[rgb]{0.608,0.761,0.902} 2017-12-28 18:02:30 & Yes                    & No                                    \\ 
	\hline
	\rowcolor[rgb]{0.663,0.816,0.557} 2017-12-29 20:00:01 & Yes                    & Yes                                   \\ 
	\hline
	\rowcolor[rgb]{0.608,0.761,0.902} 2017-12-29 20:53:30 & Yes                    & No                                    \\ 
	\hline
	\rowcolor[rgb]{0.663,0.816,0.557} 2018-01-02 09:34:36 & Yes                    & Yes                                   \\ 
	\hline
	\rowcolor[rgb]{0.663,0.816,0.557} 2018-01-08 16:58:20 & Yes                    & Yes                                   \\ 
	\hline
	\rowcolor[rgb]{0.663,0.816,0.557} 2018-01-09 20:49:00 & Yes                    & Yes                                   \\ 
	\hline
	\rowcolor[rgb]{0.608,0.761,0.902} 2018-01-12 16:20:10 & Yes                    & No                                    \\ 
	\hline
	\rowcolor[rgb]{0.608,0.761,0.902} 2018-01-13 10:29:30 & Yes                    & No                                    \\ 
	\hline
	\rowcolor[rgb]{0.608,0.761,0.902} 2018-01-19 06:49:00 & Yes                    & No                                    \\ 
	\hline
	\rowcolor[rgb]{0.608,0.761,0.902} 2018-01-24 17:04:10 & Yes                    & No                                    \\ 
	\hline
	\rowcolor[rgb]{0.608,0.761,0.902} 2018-01-29 02:57:10 & Yes                    & No                                    \\
	\hline
\end{supertabular}

\ifCLASSOPTIONcaptionsoff
  \newpage
\fi


\bibliographystyle{IEEEtran}
\bibliography{IEEEabrv,bibliography}

\begin{thebibliography}{10}
\providecommand{\url}[1]{#1}
\csname url@samestyle\endcsname
\providecommand{\newblock}{\relax}
\providecommand{\bibinfo}[2]{#2}
\providecommand{\BIBentrySTDinterwordspacing}{\spaceskip=0pt\relax}
\providecommand{\BIBentryALTinterwordstretchfactor}{4}
\providecommand{\BIBentryALTinterwordspacing}{\spaceskip=\fontdimen2\font plus
\BIBentryALTinterwordstretchfactor\fontdimen3\font minus
  \fontdimen4\font\relax}
\providecommand{\BIBforeignlanguage}[2]{{%
\expandafter\ifx\csname l@#1\endcsname\relax
\typeout{** WARNING: IEEEtran.bst: No hyphenation pattern has been}%
\typeout{** loaded for the language `#1'. Using the pattern for}%
\typeout{** the default language instead.}%
\else
\language=\csname l@#1\endcsname
\fi
#2}}
\providecommand{\BIBdecl}{\relax}
\BIBdecl

\bibitem{brodsky2019importance}
E.~E. Brodsky, ``The importance of studying small earthquakes,''
  \emph{Science}, vol. 364, no. 6442, pp. 736--737, 2019.

\bibitem{allen1978automatic}
R.~V. Allen, ``Automatic earthquake recognition and timing from single
  traces,'' \emph{Bulletin of the Seismological Society of America}, vol.~68,
  no.~5, pp. 1521--1532, 1978.

\bibitem{shelly2007non}
D.~R. Shelly, G.~C. Beroza, and S.~Ide, ``Non-volcanic tremor and low-frequency
  earthquake swarms,'' \emph{Nature}, vol. 446, no. 7133, pp. 305--307, 2007.

\bibitem{yoon2015earthquake}
C.~E. Yoon, O.~O’Reilly, K.~J. Bergen, and G.~C. Beroza, ``Earthquake
  detection through computationally efficient similarity search,''
  \emph{Science advances}, vol.~1, no.~11, p. e1501057, 2015.

\bibitem{poliannikov2018instantaneous}
O.~V. Poliannikov and M.~C. Fehler, ``Instantaneous phase-based statistical
  method for detecting seismic events with application to groningen gas field
  data,'' in \emph{SEG Technical Program Expanded Abstracts 2018}.\hskip 1em
  plus 0.5em minus 0.4em\relax Society of Exploration Geophysicists, 2018, pp.
  2907--2911.

\bibitem{mukuhira2020low}
Y.~Mukuhira, O.~V. Poliannikov, M.~C. Fehler, and H.~Moriya, ``Low-snr
  microseismic detection using direct p-wave arrival polarization,''
  \emph{Bulletin of the Seismological Society of America}, vol. 110, no.~6, pp.
  3115--3129, 2020.

\bibitem{akram2017robust}
J.~Akram, O.~Ovcharenko, and D.~Peter, ``A robust neural network-based approach
  for microseismic event detection,'' in \emph{SEG Technical Program Expanded
  Abstracts 2017}.\hskip 1em plus 0.5em minus 0.4em\relax Society of
  Exploration Geophysicists, 2017, pp. 2929--2933.

\bibitem{qu2018automatic}
S.~Qu, E.~Verschuur, and Y.~Chen, ``Automatic microseismic-event detection via
  supervised machine learning,'' in \emph{Seg technical program expanded
  abstracts 2018}.\hskip 1em plus 0.5em minus 0.4em\relax Society of
  Exploration Geophysicists, 2018, pp. 2287--2291.

\bibitem{lecun2015deep}
Y.~LeCun, Y.~Bengio, and G.~Hinton, ``Deep learning,'' \emph{nature}, vol. 521,
  no. 7553, pp. 436--444, 2015.

\bibitem{szegedy2015going}
C.~Szegedy, W.~Liu, Y.~Jia, P.~Sermanet, S.~Reed, D.~Anguelov, D.~Erhan,
  V.~Vanhoucke, and A.~Rabinovich, ``Going deeper with convolutions,'' in
  \emph{Proceedings of the IEEE conference on computer vision and pattern
  recognition}, 2015, pp. 1--9.

\bibitem{simonyan2014very}
K.~Simonyan and A.~Zisserman, ``Very deep convolutional networks for
  large-scale image recognition,'' \emph{arXiv preprint arXiv:1409.1556}, 2014.

\bibitem{kong2019machine}
Q.~Kong, D.~T. Trugman, Z.~E. Ross, M.~J. Bianco, B.~J. Meade, and P.~Gerstoft,
  ``Machine learning in seismology: Turning data into insights,''
  \emph{Seismological Research Letters}, vol.~90, no.~1, pp. 3--14, 2019.

\bibitem{perol2018convolutional}
T.~Perol, M.~Gharbi, and M.~Denolle, ``Convolutional neural network for
  earthquake detection and location,'' \emph{Science Advances}, vol.~4, no.~2,
  p. e1700578, 2018.

\bibitem{keranen2013potentially}
K.~M. Keranen, H.~M. Savage, G.~A. Abers, and E.~S. Cochran, ``Potentially
  induced earthquakes in oklahoma, usa: Links between wastewater injection and
  the 2011 mw 5.7 earthquake sequence,'' \emph{Geology}, vol.~41, no.~6, pp.
  699--702, 2013.

\bibitem{mousavi2019cred}
S.~M. Mousavi, W.~Zhu, Y.~Sheng, and G.~C. Beroza, ``Cred: A deep residual
  network of convolutional and recurrent units for earthquake signal
  detection,'' \emph{Scientific reports}, vol.~9, no.~1, pp. 1--14, 2019.

\bibitem{mousavi2020earthquake}
S.~M. Mousavi, W.~L. Ellsworth, W.~Zhu, L.~Y. Chuang, and G.~C. Beroza,
  ``Earthquake transformer—an attentive deep-learning model for simultaneous
  earthquake detection and phase picking,'' \emph{Nature communications},
  vol.~11, no.~1, pp. 1--12, 2020.

\bibitem{van2015induced}
K.~van Thienen-Visser and J.~Breunese, ``Induced seismicity of the groningen
  gas field: History and recent developments,'' \emph{The Leading Edge},
  vol.~34, no.~6, pp. 664--671, 2015.

\bibitem{reuters.USKCN1VV1KE}
\BIBentryALTinterwordspacing
B.~H. Meijer, ``Netherlands to halt groningen gas production by 2022,''
  \emph{Reuters}, 2019. [Online]. Available:
  \url{https://www.reuters.com/article/us-netherlands-gas-idUSKCN1VV1KE}
\BIBentrySTDinterwordspacing

\bibitem{dost2017development}
B.~Dost, E.~Ruigrok, and J.~Spetzler, ``Development of seismicity and
  probabilistic hazard assessment for the groningen gas field,''
  \emph{Netherlands Journal of Geosciences}, vol.~96, no.~5, pp. s235--s245,
  2017.

\bibitem{knmi1993netherlands}
KNMI, ``Netherlands seismic and acoustic network. royal netherlands
  meteorological institute (knmi). other/seismic network,'' 1993.

\bibitem{hubel1959receptive}
D.~H. Hubel and T.~N. Wiesel, ``Receptive fields of single neurones in the
  cat's striate cortex,'' \emph{The Journal of physiology}, vol. 148, no.~3,
  pp. 574--591, 1959.

\bibitem{hubel1968receptive}
------, ``Receptive fields and functional architecture of monkey striate
  cortex,'' \emph{The Journal of physiology}, vol. 195, no.~1, pp. 215--243,
  1968.

\bibitem{rumelhart1985learning}
D.~E. Rumelhart, G.~E. Hinton, and R.~J. Williams, ``Learning internal
  representations by error propagation,'' California Univ San Diego La Jolla
  Inst for Cognitive Science, Tech. Rep., 1985.

\bibitem{tensorflow2015-whitepaper}
\BIBentryALTinterwordspacing
M.~Abadi, A.~Agarwal, P.~Barham, E.~Brevdo, Z.~Chen, C.~Citro, G.~S. Corrado,
  A.~Davis, J.~Dean, M.~Devin, S.~Ghemawat, I.~Goodfellow, A.~Harp, G.~Irving,
  M.~Isard, Y.~Jia, R.~Jozefowicz, L.~Kaiser, M.~Kudlur, J.~Levenberg,
  D.~Man\'{e}, R.~Monga, S.~Moore, D.~Murray, C.~Olah, M.~Schuster, J.~Shlens,
  B.~Steiner, I.~Sutskever, K.~Talwar, P.~Tucker, V.~Vanhoucke, V.~Vasudevan,
  F.~Vi\'{e}gas, O.~Vinyals, P.~Warden, M.~Wattenberg, M.~Wicke, Y.~Yu, and
  X.~Zheng, ``{TensorFlow}: Large-scale machine learning on heterogeneous
  systems,'' 2015, software available from tensorflow.org. [Online]. Available:
  \url{http://tensorflow.org/}
\BIBentrySTDinterwordspacing

\bibitem{ross2018p}
Z.~E. Ross, M.-A. Meier, and E.~Hauksson, ``P wave arrival picking and
  first-motion polarity determination with deep learning,'' \emph{Journal of
  Geophysical Research: Solid Earth}, vol. 123, no.~6, pp. 5120--5129, 2018.

\end{thebibliography}

\end{document}